\documentclass[a4paper,11pt]{article}
\usepackage{latexsym}
\usepackage{rotating}
\usepackage{amsmath}
\author {Wen-Xiu Ma }
\title
{Integrable Couplings of Soliton Equations by Perturbations\\
I. A General Theory and Application to the KdV Hierarchy}
\date{\nonumber}
\begin{document}

\setlength{\baselineskip}{17pt}
\maketitle

\
\vskip 1cm 

\begin{abstract}
A theory for constructing integrable couplings of soliton equations
is developed by using various perturbations around solutions of perturbed 
soliton equations being analytic with respect to a small perturbation 
parameter. Multi-scale perturbations can be taken and thus higher 
dimensional integrable couplings can be presented. The theory is applied 
to the KdV soliton hierarchy. Infinitely many integrable couplings are 
constructed for each soliton equation in the KdV hierarchy, which contain 
integrable couplings possessing quadruple Hamiltonian formulations 
and two classes of hereditary recursion operators, and integrable 
couplings possessing local $2+1$ dimensional bi-Hamiltonian formulations 
and consequent $2+1$ dimensional hereditary recursion operators.
  
\end{abstract}

\vskip 1cm
\noindent {\bf Running title:} Integrable couplings by perturbations I.

\vskip 2cm 



\newcommand{\R}{\mbox{\rm I \hspace{-0.9em} R}}

\def \J {\hat{J}_N}
\def \la {\lambda}
\def \La {\Lambda}
\def \be {\beta}
\def \al{\alpha}
\def \del{\delta}
\def \Del{\Delta}
\def \al{\alpha}
\def \vare{\varepsilon}

\def \part {\partial}
\def \h1 \hat{\eta }
\def \be {\begin{equation}}
\def \ee {\end{equation}}
\def \ba {\begin{array}}
\def \ea {\end{array}}
\def\bea{\begin{eqnarray}}
\def\eea{\end{eqnarray}}

\newcommand{\eqnsection}{
   \renewcommand{\theequation}{\thesection.\arabic{equation}}
   \makeatletter
   \csname $addtoreset\endcsname
   \makeatother}
\eqnsection

\newtheorem{lemma}{Lemma}[section]
\newtheorem{theorem}{Theorem}[section]
\newtheorem{definition}{Definition}[section]

\newpage
\section{Introduction}
\setcounter{equation}{0}

Integrable couplings are a quite new interesting aspect in the field of 
soliton theory \cite{Fuchssteiner-book1993}. It originates from an 
investigation on centerless Virasoro symmetry algebras of integrable 
systems or soliton equations. The Abelian parts of those Virasoro 
symmetry algebras correspond to isospectral flows from higher order 
Lax pairs and the non-Abelian parts, to non-isospectral flows from 
non-isospectral Lax pairs \cite{Ma-JPM1992,Ma-JPAPLA199293}.
If we make a given system of soliton equations and
each time part of Lax pairs of its hierarchy to be
the first component and the second component of a new system respectively,
then such a new system will keep the same structure of 
Virasoro symmetry algebras as the old one. Therefore this can
lead to a hierarchy of integrable couplings for the original system.

Mathematically, the problem of integrable couplings may be expressed as:
{\it for a given integrable system of evolution equations $u_t=K(u)
,$ how can we  construct a non-trivial system of evolution equations
which is still integrable and includes $u_t=K(u)$ as a sub-system?}

Therefore, up to a permutation (note that we can put some components
of $u_t=K(u)$ seperately),
we actually want to construct a new bigger integrable system as follows
\begin{equation} \left\{ \begin{array} {l} u_t=K(u),\\ v_t=S(u,v),
\end{array} \right. \label{couplingsystem}\end{equation}
which should satisfy the non-triviality condition
$ \part S /\part [u] \ne 0.$
Here $[u]$ denotes a vector consisting of all derivatives of $u$ 
with respect to a space variable. For example, we  have 
$[u]=(u,u_x,u_{xx},\cdots)$ in the case of $1+1$ dimensions.
The non-triviality condition guarantees that trivial diagonal
systems with $S(u,v)=c K(v)$ are excluded, where $c $ is an arbitrary
constant.

There are two facts which have a direct relation to
the study of integrable couplings.
First, all possible methods for constructing integrable couplings
will tell us how to extend integrable systems,
from small to large and from simple to complicated,
and/or how to hunt for new integrable systems, which are probably difficult
to find in other ways. The corresponding theories may also provide useful
information for completely classifying integrable systems in whatever
dimensions. Secondly, the symmetry problem of integrable systems can be
viewed as a special case of integrable couplings. 
Strictly speaking, if a system of evolution equations $u_t=K(u)$
is integrable, then a new system (called a perturbation system)
consisting of the original system and its linearized system
\begin{equation}
\left \{ \begin{array} {l} u_t=K(u),\\ v_t=K'(u)[v],
\end{array}   \right. \nonumber
\end{equation} 
must be still integrable \cite{Fuchssteiner-book1993}.
The second part of the above new system is exactly the system that
all symmetries need to satisfy, but new system itself is 
a special integrable coupling of the original system $u_t=K(u)$.
Generally, the search for the approximate solutions
$\hat {u}_N=\sum_{i=0}^N \varepsilon ^i \eta _i,\ N \ge 1$,
of physical interest to a given system $u_t=K(u)$ 
can be cast into a study of the general standard perturbation systems
$\eta _{it}=\frac 1{i!} \frac {\part ^iK(\hat{u}_N)}{\part \varepsilon ^i},
\ 0\le i\le N$. These perturbation systems were proved to form
integrable couplings of the original system $u_t=K(u)$
\cite{LakshmananT-JMP1985,MaF-CSF1996}, the simplest case of which
is the above system associated with the symmetry problem.
This fact is also a main motivation for us to
consider the problem of integrable couplings.

However the standard perturbation systems above
are just special examples of integrable couplings.
They keep the spatial dimensions of given integrable systems invariant
and only the perturbations around solutions of
unperturbed integrable systems have been considered.
We already know \cite{Ma-Needs98} that it is possible to extend
the standard perturbation systems and to change the spatial dimensions,
in order to make more examples of integrable couplings.
The question we want to ask here is how to do generally,
or what related theory we can develop.
In this paper, we would like to provide our partial answer to this extensive
question, by establishing a theory on the multi-scale perturbation systems
of perturbed integrable systems.
An approach for extending the standard perturbation systems
and for enlarging the spatial dimensions by perturbations will be proposed.

Let us now introduce our basic 
notation and conception, some notation of which comes from
Refs. 
\cite{FuchssteinerF-PD1981,Oevel-JMP1988,Fuchssteiner-book1993,MaF-CSF1996}. 
Let $M=M(u)$ be a suitable manifold possessing a manifold variable $u$,
which is assumed to be a column vector of $q$ functions of $t\in \R$ and
$x\in \R^p$ with $t$ playing the role of time and $x$ representing
position in space. We are concerned with coupling systems
by perturbations and thus need
to introduce another bigger suitable manifold 
$\hat {M}_N=\hat {M}_N(\hat {\eta   }_N)$ possessing a manifold variable
$\hat {\eta   }_N =(\eta   _0^T,\eta   _1^T,\cdots,\eta   _N^T)^T,$ $ N\ge0$, 
where $\eta _i$, $0\le i\le N,$ are also assumed to be
column function vectors of the same dimension as $u$
and $T$ means the transpose of matrices. Assume that 
$T(M), T(\hat{M}_N)$ denote the tangent bundles on $M$ and $\hat{M}_N$,
 $T^*(M), T^*(\hat{M}_N)$ denote the cotangent bundles
 on $M$ and $\hat{M}_N$, and $C^\infty (M)$, $C^\infty (\hat{M}_N)$
denote the spaces of smooth functionals on $M$ and $ \hat{M}_N$, respectively.
Moreover let $T^r_s(M)$ be the s-times co- and r-times contravariant 
tensor bundle and $T^r_s|_u(M)$, 
the space of s-times co- and r-times contravariant 
tensors at $u\in M$. We use $X(u)$ (not $X|_u$) to denote a tensor
 of $X\in T_s^r(M)$ at $u\in M$ but sometimes we omit the point $u$ 
for brevity if there is no confusion. Note that four linear operators
$\Phi :T(M)\to T(M)$, $\Psi :T^*(M)\to T^*(M)$, $
J :T^*(M)\to T(M)$, $\Theta :T(M)\to T^*(M)$
may be identified with the second-degree tensor fields
$T_{\Phi}\in T^1_1(M),\ T_{\Psi}\in T^1_1(M),\
T_{J}\in T^2_0(M),\ T_{\Theta }\in T^0_2(M)$
by the following relations
\[\begin{array} {l} T_\Phi (u)(\al (u),K(u))=<\al (u), \Phi (u)K(u)>,\ \al 
\in T^*(M),\ K\in T(M),
\vspace {1mm}\\
 T_\Psi (u)(\al (u),K(u))=<\Psi (u)\al (u), K(u)>,\ \al 
\in T^*(M),\ K\in T(M),
\vspace {1mm}\\
T_J (u)(\al (u),\beta (u))=<\al (u), J (u)\beta (u)>,\ \al ,\beta \in T^*(M),
\vspace {1mm}\\
T_\Theta (u)(K(u),S(u))=<\Theta (u)K(u), S(u)>,\ K,S\in T(M),
 \end{array} \]
where $<\cdot,\cdot>$ denotes the duality between cotangent vectors and
tangent vectors.  

Of fundamental importance is
the conception of the Gateaux derivative, which provides a tool
to handle various tensor fields. For a tensor field $X\in T_s^r(M)$,
its Gateaux derivative at a direction $Y\in T(M)$ is defined by
\begin{equation} X'(u)[Y(u)]=\left.\frac {\part X(u+\varepsilon  
Y(u))}{\part \varepsilon  }\right|_{\varepsilon  =0}.\end{equation} 
For those operators between the tangent bundle and 
the cotangent bundle,
their Gateaux derivatives may be given similarly or 
by means of their tensor fields.
The commutator of two vector fields $K,S\in T(M)$ and the adjoint map
$\textrm{ad}_K:T(M)\to T(M)$ are commonly defined by
\begin{equation} [K,S](u)=K'(u)[S(u)]-S'(u)[K(u)],\ 
\textrm{ad}_KS=[K,S].\label{commutatorofvectorfield}\end{equation}
Note that there are some authors who use the other commutator
\[ [K,S](u)=S'(u)[K(u)]-K'(u)[S(u)].\] 
It doesn't matter of course but each type has many proponents and
hence one must be careful of plus and minus signs in reading various sources.

The conjugate operators of operators between the tangent bundle and 
the cotangent bundle are determined in terms of the duality
between cotangent vectors and tangent vectors. For instance, 
the conjugate operator $\Phi ^\dagger :T^*(M)\to T^*(M)$
of an operator $\Phi :T(M)\to T(M)$
is established by
\[ <\Phi ^\dagger (u)\al (u) ,K(u) >=<\al (u),\Phi(u) K(u)>,
\ \al \in T^*(M),\ K\in T(M).\]
If an operator $J:T^*(M)\to T(M)$ (or $\Theta :T(M)\to T^*(M)$)
plus its conjugate operator is equal to zero, then it is called 
to be skew-symmetric. 

\begin{definition} \label{def:gradientfield}
For a functional $\tilde H \in C^\infty (M)$, 
its variational derivative $\frac{\delta \tilde{H}}{\delta u}
\in T^*(M)$
is defined by 
 \[<\frac{\delta \tilde {H}}{\delta u}(u),K(u)>
=\tilde{H}'(u)[K(u)],\ K\in T(M).
\]
If for $\gamma \in T^*(M)$ there exists a functional $\tilde{H}\in
C^\infty (M)$ so that
\[ \frac {\delta \tilde{H}}{\delta u}=\gamma,\
\textrm{i.e.,}\ \tilde{H}'(u)[K(u)]=<\gamma (u)
,K(u)>,\ K\in T(M),\label{gradient}\]
then $\gamma \in T^*(M)$ is called a gradient field  with 
a potential $\tilde{H}$.
\end{definition}

A cotangent vector field 
$\gamma \in T^*(M)$ is a gradient field  if and only if  
\begin{eqnarray} && (d\gamma)(u)(K(u),S(u))\nonumber \\
&:=& <\gamma '(u)[K(u)],S(u)>-
<\gamma '(u)[S(u)],K(u)>=0,\ K,S\in T(M).
\label{defofdifferentialofcotangentvector}\end{eqnarray} 
If $\gamma \in T^*(M)$ is gradient, then its potential $\tilde{H}$
is given by 
\[ \tilde{H}(u)=\int _0^1<\gamma (\la u),u>d\lambda.\]

\begin{definition} For a linear operator 
$\Phi :T(M)\to T(M)$ and a vector field $K\in T(M)$,
the Lie derivative $L_K\Phi:T(M)\to T(M) $
of $\Phi $ with respect to $K$ is defined by
\begin{equation} 
(L_K\Phi )(u)S(u)=\Phi '(u)[K(u)]S(u)-K'(u)[\Phi (u)S(u)]
+\Phi (u)K'(u)[S(u)],\  S\in T(M).\end{equation}
\end{definition}

An equivalent form of the Lie derivative is
\begin{equation} (L_K\Phi)(u)S(u)=\Phi (u)[K(u),S(u)]-[K(u),\Phi (u) S(u)],
\label{anotherformforLiederivativeofPhi}\end{equation}
where $\Phi :T(M)\to T(M)$, $K,S\in T(M)$, and the commutator
$[\cdot,\cdot]$ is defined by (\ref{commutatorofvectorfield}).

\begin{definition}
A linear operator $\Phi :T(M)\to T(M)$ is called a recursion operator 
of $u_t=K(u),\ K\in T(M),$ if for all $S\in T(M)$ and $u\in M$,
we have
\begin{equation} \frac {\part \Phi(u)}{\part t}S(u)+
\Phi '(u)[K(u)]S(u)-K'(u)[\Phi (u)S(u)]+\Phi (u)K'(u)[S(u)]=0.\end{equation} 
\label{def:recursionoperator}
\end{definition}

Obviously a recursion operator $\Phi :T(M)\to T(M)$ of 
a system $u_t=K(u),\ K\in T(M),$ 
transforms a symmetry into another symmetry of the same system $u_t=K(u)$.
Therefore it is very useful
in constructing the corresponding
symmetry algebra of a given system
and its existence is regarded as an
important characterizing property for integrability of the system
under study.
 
\begin{definition}
\label{def:hereditaryoperator}
A linear operator $\Phi :T(M)\to T(M)$ is called a hereditary
operator or to be hereditary
\cite{Fuchssteiner-NATMA1979}, if the following equality holds
\begin{equation}\begin{array} {l} \Phi '(u)[\Phi(u) K(u)]S(u)-\Phi(u) 
\Phi '(u)[K(u)]S(u)
\vspace{1mm}\\ \ 
-\Phi '(u)[\Phi (u)S(u)
]K(u)
+\Phi(u) \Phi '(u)[S(u)]K(u)=0\end{array}\label{hereditarydefinition}
 \end{equation} 
for all  vector fields $K,S\in T(M)$.
\end{definition}

For a linear operator $\Phi :T(M)\to T(M)$, 
The above equality (\ref{hereditarydefinition})
can be replaced with either of the following equalities:
\[\begin{array}{l} (L_{\Phi K})(u)\Phi (u) =\Phi (u)(L_K\Phi)(u) ,
\ K\in T(M),\vspace{2mm} \\
\Phi ^2(u)[K(u),S(u)]+[\Phi (u) K(u),\Phi (u)S(u)]
\vspace{2mm} \\
\quad -\Phi(u)\{[K(u),\Phi (u)S(u)]+[\Phi (u)K(u),S(u)]\}=0,\ K,S\in T(M).
\end{array}\]
It follows directly from (\ref{anotherformforLiederivativeofPhi}) that
these two equalities are equivalent to each other.
We point out that hereditary operators have two remarkable properties.
First, if $\Phi:T(M)\to T(M)$ is
hereditary and $L_K\Phi =0, \ K\in T(M)$,
then we have $[\Phi ^mK,\Phi ^nK]=0,\ m,n\ge 0$ (see, for example, 
\cite{Fuchssteiner-NATMA1979,Fokas-SAM1987,Gu-book1995}). Therefore,  
when a system $u_t=K(u),\ K\in T(M),$ possesses
a time-independent hereditary recursion operator $\Phi:T(M)\to T(M) $,
a hierarchy of vector fields
 $\Phi ^nK, \ n\ge 0$, are all symmetries and commute with each other.
Secondly, if the conjugate operator $\Psi =\Phi ^\dagger $ of 
a hereditary operator $\Phi :T(M)\to T(M)$ maps a gradient field  
$\gamma \in T^*(M)$ into another gradient field, then $\Psi ^n\gamma ,\
 n\ge0$, are all gradient fields (see, for example, \cite{Olver-book1986}).

\begin{definition} 
A linear skew-symmetric operator
$J:T^*(M)\to T(M)$ is called a Hamiltonian operator or to be Hamiltonian,
if for all $\alpha , \beta , \gamma \in T^*(M)$, we have
\begin{equation} <\alpha ,J'(u)[J(u)\beta ]\gamma >
+\textrm{cycle}(\alpha , \beta , \gamma )=0.\end{equation}
Its Poisson bracket is defined by 
\begin{equation} \{\tilde{H}_1,\tilde{H}_2\}_J(u)
=<\frac {\delta \tilde{H}_1}{\delta u}(u),J(u)
\frac {\delta \tilde{H}_2}{\delta u}(u)>,\end{equation}
where $\tilde{H}_1,\, \tilde{H}_2 \in C^\infty (M)$.
A pair of operators $J,M:T^*(M)\to T(M)$ is called a Hamiltonian
pair, if $cJ+d M$ is always Hamiltonian for any constants $c,d $.
\end{definition}

When $J:T^*(M)\to T(M)$ is Hamiltonian, we have \cite{GelfandD-FAA1979}
\[ J(u)\frac {\delta \{\tilde {H}_1,\tilde{H}_2\}_J }{\delta u}(u)
=[J(u)\frac {\delta \tilde{H}_1}{\delta u}(u),
J(u)\frac {\delta \tilde{H}_2}{\delta u}(u)], \]
where $\tilde{H}_1,\tilde{H}_2 \in C^\infty (M)$.
This implies that the operator $J\frac{\delta }{\delta u}$
is a Lie homomorphism from the Poisson algebra to the vector field algebra.
Moreover if $J,M:T^*(M)\to T(M)$ is a Hamiltonian pair
and $J$ is invertible, then $\Phi =MJ^{-1}:T(M)\to T(M)$ defines a 
hereditary operator 
\cite{GelfandD-FAA1979,FuchssteinerF-PD1981}.

\begin{definition}A linear skew-symmetric
operator $\Theta :T(M)\to T^*(M)$ is called a
symplectic operator or to be symplectic, if for all $K,S,T\in T(M)$, we have
\begin{equation} <K(u),\Theta '(u)[S(u)]T(u)>+\textrm{cycle}(K,S,T)=0.
\end{equation}
\end{definition}

If $\Theta :T(M)\to T^*(M)$ is a symplectic operator, then its
second-degree tensor field $T_\Theta \in T_2^0(M)$ can be expressed as
\[ T_\Theta =d\gamma \ \textrm{with}\  
 <\gamma (u),K(u)>=\int _0^1<\Theta (\la u)\la u, K(u)>\, d\lambda ,
\ K\in T(M),\]
where $d\gamma $ is determined by (\ref{defofdifferentialofcotangentvector}).
It is not difficult to verify that the inverse of a symplectic operator
is Hamiltonian if it exists and vice versa. We also mention that 
Hamiltonian and symplectic operators can be defined only 
in terms of Dirac structures \cite{Dorfman-book1993}.

\begin{definition}
A system of evolution equations $u_t=K(u),\ K\in T(M),$ is called
a Hamiltonian system or to be Hamiltonian, if there exists a functional
$\tilde{H}\in C^\infty (M)$ so that
\begin{equation} u_t=K(u)=J(u)\frac{\delta \tilde{H}}
{\delta u}(u).\end{equation}
It is called a bi-Hamiltonian system, if there exist two functionals
$\tilde{H}_1,\tilde{H}_2\in C^\infty (M)$ and a Hamiltonian pair
$J,M:T^*(M)\to T(M)$ so that
\begin{equation}u_t=K(u)=J(u)\frac{\delta \tilde{H}_1}{\delta u}(u)
=M(u)\frac{\delta \tilde{H}_2}{\delta u}(u).\end{equation} 
\end{definition}

There is the other kind of Hamiltonian systems, which 
can be defined by symplectic operators. However, the above definition 
has more advantages in handling symmetries and conserved functionals.
For a Hamiltonian system $u_t=J(u)\frac{\delta \tilde{H}}
{\delta u}(u)$, the linear operator
$J\frac{\delta }{\delta u}$ maps a conserved
functional into a symmetry. For a bi-Hamiltonian system, 
we will be able to recursively construct infinitely many commuting
symmetries and conserved functionals for the system, if either of two
Hamiltonian operators is invertible \cite{Magri-JMP1978}. 

In what follows, we would like to develop a theory for constructing
integrable couplings of soliton equations, by analyzing integrable 
properties of the perturbation systems resulted from perturbed soliton 
equations by multi-scale perturbations. The paper is organized as follows.
In Section 2, we first establish general explicit structures of
hereditary operators, Hamiltonian operators and symplectic operators 
under the multi-scale perturbations. We will go on to show that 
the perturbations preserves complete integrability, by establishing 
various integrable properties of the resulting perturbation systems, such as
hereditary recursion operator structures, Virasoro symmetry algebras,
Lax representations, zero curvature representations, Hamiltonian formulations 
and so on. In Section 3, the whole theory will be applied to the KdV equations
as illustrative examples. This leads to infinitely many integrable couplings 
of the KdV equations, which include Hamiltonian integrable couplings
possessing two different hereditary recursion operators and local
bi-Hamiltonian integrable couplings in $2+1$ dimensions. Finally,
some concluding remarks are given in Section 5.

\section{Integrable couplings by perturbations}
\setcounter{equation}{0}

\subsection{Triangular systems by perturbations}

Let us take a perturbation series for any $N\ge 0$ and $r\ge 0$:
\begin{equation}
\hat{  u}_N=\sum_{i=0}^N \varepsilon  ^i\eta _i , \
\eta _i=\eta _i(y_0, y_1, y_2,\cdots,y_r,t),\
y_i=\varepsilon ^i x, \ t\in \R,\ x\in \R ^p,\ 0\le i\le N,
\label{multiplescaleperturbationseries}
\end{equation}
where $\varepsilon $ is a perturbation parameter
and $\eta _i,\ 0\le i\le N,$ are assumed to be column vectors of 
$q$ dimensions as before. When $r\ge 1$, 
(\ref{multiplescaleperturbationseries}) is really a multi-scale 
perturbation series. We fix a perturbed vector field 
$K=K(\varepsilon)\in T(M)$ which is required to be  
analytic with respect to $\varepsilon$. Let us introduce
\begin{equation} K^{(i)}=K^{(i)}(\hat{\eta }_N)=
\bigr( K(u,\varepsilon )\bigl)^{(i)}(\hat {\eta   }_N)= \frac 1 {i!}
\left.\frac {\part ^i K(\hat {u}_N,\varepsilon)}{\part \varepsilon  ^i}
\right | _{\varepsilon  =0},  \
 0\le i\le N,
\end{equation}
where $\hat {\eta   }_N =(\eta   _0^T,\eta   _1^T,\cdots,\eta   _N^T)^T$
as before, and then define the $N$-th order perturbation vector field on 
$\hat {M}_N$:
\begin{equation} (\textrm{per}_NK)(\hat {\eta }_N)=
\hat {K}_N(\hat {\eta }_N)=(K^{(0)T}(\hat {\eta }_N),K^{(1)T}(\hat {\eta }_N)
,\cdots,K^{(N)T}(\hat {\eta }_N))^T. \label{perturbationvectorfield}
\end{equation} 
Here the vector fields on $M$ are viewed as column vectors of
$q$ dimensions, and the vector fields on $\hat{M}_N$, column vectors
of $q(N+1)$ dimensions, as they are normally handled. 
Since we have
\[ \left.\frac {\part ^i K(\hat{u}_i,\varepsilon)}{\part \varepsilon ^i}
\right|_{\varepsilon =0}=
\left.\frac {\part ^i K(\hat{u}_j,\varepsilon)}{\part \varepsilon ^i}
\right|_{\varepsilon =0}, \
\hat{u}_i=\sum_{k=0}^i\varepsilon ^k
\eta _k,\ \hat{u}_j=\sum_{k=0}^j\varepsilon ^k
\eta _k,\ 0\le i\le j\le N, \]
it is easy to find that
\begin{equation}\hat {K}_N(\hat{\eta}_N) =(K^{(0)T}(\hat{\eta}_0),K^{(1)T}
(\hat{\eta}_1),\cdots,K^{(N)T}(\hat{\eta}_N))^T,
\ \hat{\eta }_i=(\eta _0^T,\eta _1^T,\cdots, \eta _i^T)^T,\ 0\le i\le N.
\label{triangularpropertyofpvf}
\end{equation}
Thus the perturbation vector field $\textrm{per}_NK=\hat{K}_N
\in T(\hat{M}_N)$ has a specific property that
the $i$-th component depends only on
$\eta _0,\eta _1,\cdots , \eta _i$,
not on any $\eta _j,\ j>i$.

Let us now consider a system of perturbed evolution equations 
\begin{equation}
u_{t}=K(u,\varepsilon ),\ K=K(\varepsilon)\in T(M),
\label{initialperturbedsystem}\end{equation}
where $K(\varepsilon)$ is assumed to be analytic with respect to 
$\varepsilon$, as an initial system that we start from.   
It is obvious that the following perturbed system
\begin{equation}
\hat {u}_{Nt}=K({\hat{u}_N},\varepsilon )
+\textrm{o}(\varepsilon  ^N)\ \ \textrm{or}\ \ 
\hat {u}_{Nt} \equiv K({\hat{u}_N},\varepsilon )
\  \pmod{\varepsilon  ^N},\end{equation}
leads equivalently to a bigger system of evolution equations
\begin{equation}
 \hat {\eta   }_{Nt}=\hat {K}_{N}({\hat{\eta}_N} ),\ \  \textrm{i.e.,}\ \ 
\eta   _{it}=\left.
\frac1 {i!} \frac {\part ^iK({\hat{u}_N},\varepsilon )}
{\part \varepsilon  ^i}\right|_{\varepsilon  =0},\ 0\le i\le N,
\label{multiplescaleperturbationsystem}
\end{equation}
where $\hat {u}_N$ is defined by (\ref{multiplescaleperturbationseries}).
Conversely, a solution $\hat{\eta }_N$ of the bigger system 
(\ref{multiplescaleperturbationsystem}) gives rise to an approximate solution
$\hat{u}_N$ of the initial system (\ref{initialperturbedsystem}) 
to a precision $\textrm{o}(\varepsilon ^N)$.
The resulting bigger system (\ref{multiplescaleperturbationsystem})
is called an $N$-th order perturbation system of 
the initial perturbed system (\ref{initialperturbedsystem}),
and it is a triangular system, owing to (\ref{triangularpropertyofpvf}).
We will analyze its integrable properties by exposing structures 
of other perturbation objects.
                                                  
\subsection{Symmetry problem}

Let us shed more right on an remarkable relation between the symmetry problem 
and integrable couplings. Assume that a system $u_t=K(u),\ K\in T(M),$ is
given. Then its linearized system reads as $v_t=K'(u)[v]$. What the symmetry
problem requires to do is to find vector fields $S\in T(M)$ which satisfy
this linearized system, i.e., $(S(u))_t=K'(u)[S(u)]$ when $u_t=K(u)$.
Therefore $(u^T,(S(u))^T)$ solves the following coupling system
\begin{equation}\left\{ \begin{array} {l}  u_t=K(u), \vspace{2mm} \\ 
v_t=K'(u)[v], \end{array} \right. \label{symmetrysystem} \end{equation} 
if $S\in T(M)$ is found to be a symmetry of $u_t=K(u)$. 
This system (\ref{symmetrysystem}) has been carefully considered upon
introducing the perturbation bundle \cite{Fuchssteiner-book1993}. 
It is the first-order standard perturbation system of $u_t=K(u)$,
introduced in Ref. \cite{MaF-CSF1996}. Since it keeps complete integrability,
it provides us with an integrable coupling of the original system $u_t=K(u)$.
Therefore the symmetry problem is viewed as a sub-case of general
integrable couplings.

The commutator of the vector fields of the form $(K(u),A(u)v)^T$ with
$A(u)$ being linear has a nice structure:
\[ \Bigl[ \Bigl( \begin{array}{c} K(u)\vspace{2mm}\\ A(u)v \end{array}\Bigr),
\Bigl( \begin{array}{c} S(u)\vspace{2mm}\\ B(u)v \end{array}\Bigr)\Bigr]=
\Bigl( \begin{array}{c} [K(u),S(u)]\vspace{2mm}\\
\lbrack\!\lbrack A(u), B(u) \rbrack\!\rbrack v \end{array}\Bigr), \]
where the commutator $[K(u),S(u)]$ is given by (\ref{commutatorofvectorfield})
and the commutator $\lbrack\!\lbrack A(u), B(u) \rbrack\!\rbrack$ 
of two linear operators $A(u),B(u)$ is defined by
\[ \lbrack\!\lbrack A(u), B(u) \rbrack\!\rbrack
=A'(u)[S(u)]-B'(u)[K(u)]+A(u)B(u)-B(u)A(u), \]
which was used to analyze algebraic structures of Lax operators in
\cite{Ma-JPA1992}. Moreover for linearized operators, we can have 
\begin{equation} 
\lbrack\!\lbrack K'(u),S'(u) \rbrack\!\rbrack =T'(u),\ T=[K,S],\ K,S\in T(M),
\label{lieproductoflinearizeoperators}
\end{equation}
which will be shown later on.

\subsection{Candidates for integrable couplings}

Let us illustrate the idea of how to construct candidates 
for integrable couplings by perturbations. 
Assume that an unperturbed system is given by 
\begin{equation}u_t=K(u),\ K\in T(M),\label{unperturbedsystem}\end{equation}
and we want to construct its integrable couplings. To this end,
let us choose a simple perturbed system 
\begin{equation}u_t=K(u)+\varepsilon K(u),
\label{firstorderperturbedinitialsystem}\end{equation}
which is analytic with respect to $\varepsilon $ of course,
as an initial system. Obviously this system doesn't change integrable 
properties of the original system (\ref{unperturbedsystem}). 
In practice, we can have lots of choices of such perturbed systems.
For example, if the system (\ref{unperturbedsystem})
has a symmetry $S\in T(M)$, then we can choose either
$u_t=K(u)+\varepsilon S(u)$ or $u_t=K(u)+\varepsilon ^2S(u)$  
as another initial perturbed system. According to the definition
of the perturbation systems in (\ref{multiplescaleperturbationsystem}),
the first-order perturbation system of the above perturbed system
(\ref{firstorderperturbedinitialsystem}) reads as
\begin{equation}\left \{ \begin{array}{l} \eta _{0t}=K(\eta _0),\vspace{2mm}\\
\eta _{1t}=K'(\eta _0)[\eta _1]+K(\eta _0). \end{array}\right.  
\label{nonstandardfirstorderperturbationsystem}\end{equation}
This coupling system is a candidate that we want to 
construct for getting integrable couplings of the original system 
$u_t=K(u)$. In fact, we will verify that the perturbation defined by 
(\ref{multiplescaleperturbationseries}) preserves complete integrability.  
Therefore the above coupling system 
(\ref{nonstandardfirstorderperturbationsystem}) is an integrable coupling 
of the original system $u_t=K(u)$, provided that 
$u_t=K(u)$ itself is integrable. The realization of more integrable couplings,
such as local $2+1$ dimensional bi-Hamiltonian systems, can be found from 
an application to the KdV hierarchy in the next section.

\subsection{Structures of perturbation operators}

Rather than working with concrete examples, we would like to 
establish general structures for three kinds of perturbation operators.
The following three theorems will show us how to construct them explicitly. 
For the proof of the theorems, we first need to prove a basic result 
about the Gateaux derivative of the perturbation tensor fields.

\begin{lemma} \label{lemma:gateaxderivativeofptf}
Let $X=X(\varepsilon )\in T^r_s(M)$ be analytic with respect to $\varepsilon$
and assume that the vector 
field $\bar {S}_N=(S_0^T,S_1^T,\cdots,S_N^T)^T\in T(\hat{M}_N)$,
where all sub-vectors $S_i,\ 0\le i\le N$, are of the same dimension. Then
the following equalities hold: 
\begin{equation}
\Bigl (\left.\frac {\part ^iX 
(\hat{u}_N,\varepsilon )}{\part \varepsilon  ^i}\right|_{\varepsilon  =0}
\Bigr )'(\hat {\eta   }_N)[\bar S_N]=\left.
\frac{\part ^i}{\part \varepsilon  ^i}\right|_{\varepsilon   =0
}X'(\hat {u}_N,\varepsilon )\bigl[\sum_{
j=0}^N\varepsilon  ^jS_j\bigr],\ 0\le i\le N.
\label{gateaxderivativeofptf}
\end{equation}
\end{lemma}
{\bf Remark:} Note that in (\ref{gateaxderivativeofptf}), we have adopted
the notation
\begin{equation}X'(\hat{u}_N,\varepsilon)[K(u)]
=(X(\hat{u}_N,\varepsilon))'(\hat{u}_N) [K(u)],\ X=X(\varepsilon)
\in T^r_s(M),\ K\in T(M), \end{equation}
in order to save space. The same notation will be used in
the remainder of the paper.

\noindent {\bf Proof:}
Let us first observe Taylor series
\[ X(\hat {u}_N,\varepsilon)=\sum_{i=0}^N\frac {\varepsilon  ^i}{i!}
\left.\frac{\part ^i X(\hat {u}_N,\varepsilon)}{\part \varepsilon  ^i}
\right  |_{\varepsilon  =0}+\textrm{o}(\varepsilon  ^N).\]
It follows that
\[(X(\hat{u}_N,\varepsilon))'({\hat{\eta}_N} )[\bar{S}_N]
=\sum_{i=0}^N\frac {\varepsilon  ^i}{i!}
\Bigl(\left.\frac {\part ^iX(\hat{u}_N,\varepsilon)}{\part \varepsilon  
^i}\right|_{\varepsilon  =0}\Bigr)'({\hat{\eta}_N} )[
\bar{S}_N]+\textrm{o}(\varepsilon  ^N).\]
Secondly, we can compute that 
\[(X(\hat{u}_N,\varepsilon))'({\hat{\eta}_N} )[\bar{S}_N]=
\left.\frac {\part }{\part \delta}\right|_{\delta =0}X(\hat {u}_N+\delta
\sum_{j=0}^N\varepsilon  ^jS_j,\varepsilon)=
X'(\hat {u}_N ,\varepsilon)\Bigl [ \sum_{j=0}^N\varepsilon  ^j S_j\Bigr].\]
A combination of the above two equalities
leads to the required equalities in
(\ref{gateaxderivativeofptf}), again according to Taylor series.
The proof is completed. 
$\vrule width 1mm height 3mm depth 0mm$

\begin{theorem}  \label{thm:perturbationhereditaryoperator}
If the operator $\Phi =\Phi (\varepsilon):T(M)\to T(M)$ being analytic with
respect to $\varepsilon$ is hereditary, then the following operator
${\hat{\Phi}_N}: T(\hat{M}_N)\to T(\hat{M}_N)$ defined by
\begin{eqnarray}
&&(\textrm{per}_N\Phi)(\hat{\eta   }_N)= {\hat{\Phi}_N}({\hat{\eta}_N} )
\nonumber
\\&=&\left[ \bigl({\hat{\Phi}_N}({\hat{\eta}_N} )\bigr)_{ij}
\right]_{i,j=0,1,\cdots,N}
=\left[ \frac 1 {(i-j)!}\left.\frac {\part ^{i-j}
\Phi ({\hat{u}_N},\varepsilon )}{\part \varepsilon  ^{i-j}}
\right|_{\varepsilon  =0}
 \right]_{q(N+1)\times q(N+1)}\nonumber
\\ &= & \left[ \begin{array}{cccc}
\Phi (\hat{u}_N,\varepsilon)|_{\varepsilon=0}& & &0\vspace{1mm}\\ 
\frac {1}{1!}\left.\frac {\part \Phi ({\hat{u}_N},\varepsilon )}
{\part \varepsilon  }\right| _{\varepsilon  =0}&
\Phi (\hat{u}_N,\varepsilon)|_{\varepsilon=0}
 & &\vspace{1mm} \\ \vdots & \ddots & \ddots &\vspace{1mm} \\
\frac {1}{N!}\left.\frac {\part ^N \Phi ({\hat{u}_N},\varepsilon )}
{\part \varepsilon  ^N } \right|_{\varepsilon  =0}&\cdots &
\frac {1}{1!}\left.\frac {\part \Phi ({\hat{u}_N},\varepsilon )}
{\part \varepsilon  }\right|_{\varepsilon  =0}&
\Phi (\hat{u}_N,\varepsilon)|_{\varepsilon=0}
\end{array}
\right]\label{newhere}
\end{eqnarray}
is also hereditary, where $\hat{u}_N$ is a perturbation series
defined by (\ref{multiplescaleperturbationseries}).
\end{theorem}
{\bf Proof:} Let $\bar{ K}_N=(K_0^T,K_1^T,\cdots,K_N^T)^T,$ $
\bar{ S}_N=(S_0^T,S_1^T,\cdots,S_N^T)^T\in T(\hat{M}_N)$,
where the sub-vectors $K_i,S_i,\ 0\le i\le N$, are of the same dimension.
Since $\hat {\Phi}_N(\hat {\eta }_N)$ is obviously linear,
we only need to prove that
\begin{eqnarray}
&&{\hat{\Phi}_N}'({\hat{\eta}_N} )[{\hat{\Phi}_N}({\hat{\eta}_N} )\bar{ K}_N]
\bar{ S}_N-{\hat{\Phi}_N}({\hat{\eta}_N} ){\hat{\Phi}_N}'({\hat{\eta}_N} )
[\bar{ K}_N]\bar{ S}_N\nonumber\\
 &&-{\hat{\Phi}_N}'({\hat{\eta}_N} )[{\hat{\Phi}_N}({\hat{\eta}_N} )
 \bar{ S}_N]\bar{ K}_N+{\hat{\Phi}_N}({\hat{\eta}_N} )
 {\hat{\Phi}_N}'({\hat{\eta}_N} ) [\bar{ S}_N]\bar{ K}_N=0, \label{heequiv}
\end{eqnarray}
according to Definition \ref{def:hereditaryoperator}.
In what follows, we are going to prove this equality.

First, we immediately obtain the $i$-th element of the vector field
${\hat{\Phi}_N}({\hat{\eta}_N} )\bar{K}_N $ and the element in the $(i,j)$
position of the matrix ${\hat{\Phi}_N}'({\hat{\eta}_N} )[\bar{K}_N]$:
 \begin{eqnarray} ({\hat{\Phi}_N}({\hat{\eta}_N} )\bar{K}_N)_i=\sum_{j=0}^i
\frac1 {(i-j)!}\left.\frac {\part ^{i-j}\Phi ({\hat{u}_N},\varepsilon )}
{\part \varepsilon  ^{i-j}}
 \right|_{\varepsilon  =0} K_j,\ 0\le i\le N,&&\nonumber\\
\bigl({\hat{\Phi}_N}'({\hat{\eta}_N} )[\bar{K}_N]\bigr)_{ij}=
\frac1 {(i-j)!}\left(\left.
\frac {\part ^{i-j}\Phi (\hat {u}_N,\varepsilon)}{\part \varepsilon  ^{i-j}}
 \right|_{\varepsilon  =0}\right)'({\hat{\eta}_N} )[\bar{K}_N]&&\nonumber\\
\qquad  =\frac1 {(i-j)!}\left.\frac {\part ^{i-j}}{\part \varepsilon  ^{i-j}}
 \right|_{\varepsilon  =0}\Bigl(\Phi '({\hat{u}_N},\varepsilon )\Bigl[
\sum_{k=0}^N\varepsilon  ^kK_k\Bigr]\Bigr),\ 0\le i,j\le N , 
&&\nonumber\end{eqnarray}
 the last equality of which follows from Lemma
 \ref{lemma:gateaxderivativeofptf}.

Now we can compute the $i$-th element of ${\hat{\Phi}_N}'({\hat{\eta}_N} )
[{\hat{\Phi}_N}({\hat{\eta}_N} )\bar{K}_N]\bar{S}_N$ as follows:
\begin{eqnarray}
& &({\hat{\Phi}_N}'({\hat{\eta}_N} )
[{\hat{\Phi}_N}({\hat{\eta}_N} )\bar{K}_N]\bar{S}_N)_i\nonumber\\&=&
\sum_{j=0}^i
\frac1 {(i-j)!}\left.\frac {\part ^{i-j}}{\part \varepsilon  ^{i-j}}
 \right|_{\varepsilon  =0}
\Phi '({\hat{u}_N},\varepsilon )\Bigl[
\sum_{k=0}^N\varepsilon  ^k \sum_{l=0}^k 
\frac1 {(k-l)!}\left.\frac {\part ^{k-l}\Phi ({\hat{u}_N},\varepsilon )}
{\part \varepsilon  ^{k-l}} \right|_{\varepsilon  =0}K_l
\Bigr]S_j\nonumber\\
& = &\sum_{j=0}^i
\frac1 {(i-j)!}\left.\frac {\part ^{i-j}}{\part \varepsilon  ^{i-j}}
 \right|_{\varepsilon  =0} \Phi '({\hat{u}_N},\varepsilon )\Bigl[
\sum_{l=0}^N\varepsilon  ^l \sum_{k=l}^N 
\frac{\varepsilon  ^{k-l}} {(k-l)!}\left.\frac 
{\part ^{k-l}\Phi ({\hat{u}_N},\varepsilon )}{\part \varepsilon  ^{k-l}}
 \right|_{\varepsilon  =0}K_l
\Bigr]S_j\nonumber\\
& =&\sum_{j=0}^i
\frac1 {(i-j)!}\left.\frac {\part ^{i-j}}{\part \varepsilon  ^{i-j}}
 \right|_{\varepsilon  =0}\Phi '({\hat{u}_N},\varepsilon )\Bigl[
\sum_{l=0}^N\varepsilon  ^l \bigl(\Phi ({\hat{u}_N},\varepsilon )
+\textrm{o}(\varepsilon  ^{N-l})\bigr)K_l
\Bigr]S_j\nonumber\\
& =&\sum_{j=0}^i
\frac1 {(i-j)!}\left.\frac {\part ^{i-j}}{\part \varepsilon  ^{i-j}}
 \right|_{\varepsilon  =0}\Phi '({\hat{u}_N},\varepsilon )\Bigl[
\sum_{l=0}^N\varepsilon  ^l \Phi ({\hat{u}_N},\varepsilon )K_l
\Bigr]S_j\nonumber\\
&=&\sum_{0\le j+l\le i}
\frac1 {(i-j-l)!}\left.\frac {\part ^{i-j-l}}{\part \varepsilon  ^{i-j-l}}
 \right|_{\varepsilon  =0}\Phi '({\hat{u}_N},\varepsilon )
\Bigl[\Phi ({\hat{u}_N},\varepsilon )K_l\Bigr]S_j,\
0\le i\le N.\nonumber
\end{eqnarray}
On the other hand, we can compute the $i$-th element of
${\hat{\Phi}_N}({\hat{\eta}_N} ){\hat{\Phi}_N}'({\hat{\eta}_N} )
[\bar {K}_N]\bar {S}_N$ as follows:
\begin{eqnarray}
&&\bigl({\hat{\Phi}_N}({\hat{\eta}_N} ){\hat{\Phi}_N}'
({\hat{\eta}_N} )[\bar {K}_N]\bar {S}_N\bigr)_i\nonumber
\\&=& \sum_{j=0}^i\frac1 {(i-j)!}\left.
\frac {\part ^{i-j}\Phi ({\hat{u}_N},\varepsilon )}
{\part \varepsilon  ^{i-j}} \right|_{\varepsilon  =0}\sum_{k=0}^j
\frac1 {(j-k)!}\left.\frac {\part ^{j-k}}{\part \varepsilon  ^{j-k}}
 \right|_{\varepsilon  =0}\Phi '({\hat{u}_N},\varepsilon )
\Bigl[\sum_{l=0}^N\varepsilon  ^lK_l\Bigr]S_k\nonumber\\
& =&\sum_{j=0}^i\frac1 {(i-j)!}\left.
\frac {\part ^{i-j}\Phi ({\hat{u}_N},\varepsilon )}
{\part \varepsilon  ^{i-j}} \right|_{\varepsilon  =0}
\sum_{ k=0}^j\sum_{l=0}^{j-k}
\frac1 {(j-k-l)!}\left.\frac {\part ^{j-k-l}}{\part \varepsilon  ^{j-k-l}}
 \right|_{\varepsilon  =0}\Phi '({\hat{u}_N} ,\varepsilon)[K_l]S_k\nonumber\\
& =&\sum_{k=0}^i\sum_{j=k}^i\sum_{l=0}^{j-k}
\frac1 {(i-j)!(j-k-l)!}\left.\frac {\part ^{i-j}\Phi ({\hat{u}_N},\varepsilon )
}{\part \varepsilon  ^{i-j}} \right|_{\varepsilon  =0}
\left.\frac {\part ^{j-k-l}}{\part \varepsilon  ^{j-k-l}}
 \right|_{\varepsilon  =0}\Phi '({\hat{u}_N} ,\varepsilon)[K_l]S_k\nonumber\\
& =&\sum_{k=0}^i\sum_{l=0}^{i-k}\sum_{j=k+l}^i
\frac1 {(i-j)!(j-k-l)!}\left.\frac {\part ^{i-j}\Phi ({\hat{u}_N},\varepsilon )
}{\part \varepsilon  ^{i-j}} \right|_{\varepsilon  =0}\left.
\frac {\part ^{j-k-l}}{\part \varepsilon  ^{j-k-l}}
 \right|_{\varepsilon  =0}\Phi '({\hat{u}_N} ,\varepsilon)[K_l]S_k\nonumber\\
& =&\sum_{k=0}^i\sum_{l=0}^{i-k} 
\frac1 {(i-k-l)!}\left.\frac {\part ^{i-k-l}}{\part \varepsilon  ^{i-k-l}}
  \right|_{\varepsilon  =0}\Bigl(\Phi ({\hat{u}_N},\varepsilon )
\Phi '({\hat{u}_N},\varepsilon ) [K_l]S_k\Bigr)\nonumber\\
& =&\sum_{0\le k+l\le i}
\frac1 {(i-k-l)!}\left.\frac {\part ^{i-k-l}}{\part \varepsilon  ^{i-k-l}}
  \right|_{\varepsilon  =0}\Bigl(\Phi ({\hat{u}_N},\varepsilon )
\Phi '({\hat{u}_N},\varepsilon ) [K_l]S_k\Bigr),\ 0\le i\le N.\nonumber
\end{eqnarray}              
Therefore, it follows from the hereditary property of $\Phi(u,\varepsilon)$
that each element in the left-hand side of
(\ref{heequiv}) is equal to zero, which means that (\ref{heequiv}) is true. 
The proof is completed.
$\vrule width 1mm height 3mm depth 0mm$

\begin{theorem} \label{thm:perturbationHamiltonianoperator}
If the operator $J=J(\varepsilon) :T^*(M)\to T(M)$ being analytic
with respect to $\varepsilon$ is Hamiltonian,
 then the following operator
 ${\hat{J}_N}:T^*(\hat{M}_N)\to T(\hat{M}_N)$ defined by
\begin{eqnarray}&&(\textrm{per}_NJ)(\hat{\eta   }_N)=
{\hat{J}_N}({\hat{\eta}_N} )\nonumber
\\
&=&\left[ \bigl({\hat{J}_N}({\hat{\eta}_N} )\bigr)_{ij}\right]_{i,j=0,1,
\cdots,N} =\left[
\frac 1 {(i+j-N)!}\left.\frac {\part ^{i+j-N}J ({\hat{u}_N} ,\varepsilon)}
{\part \varepsilon  ^{i+j-N}}
\right|_{\varepsilon  =0}
 \right]_{q(N+1)\times q(N+1)}\nonumber
\\
 &= &
\left[ \begin{array}{cccc}
0& & & J (\hat{u}_N,\varepsilon)|_{\varepsilon=0}
\\ & & J (\hat{u}_N,\varepsilon)|_{\varepsilon=0}&
\frac 1{1!}\left.
\frac {\part  J ({\hat{u}_N} ,\varepsilon )}{\part \varepsilon  }
\right|_{\varepsilon  =0} \\
 &\begin{turn}{45}\vdots\end{turn} &\begin{turn}{45}\vdots\end{turn}& 
\vdots\\ J (\hat{u}_N,\varepsilon)|_{\varepsilon=0}&\frac 1{1!}\left.
\frac {\part  J ({\hat{u}_N} ,\varepsilon )}{\part \varepsilon  }
\right|_{\varepsilon  =0}&\cdots&
\frac {1}{N!}\left.\frac {\part ^N J ({\hat{u}_N} ,\varepsilon )}
{\part \varepsilon  ^N }
\right|_{\varepsilon  =0} 
\end{array}
\right] \label{msperHamiltonianoperator}
\end{eqnarray}is also Hamiltonian,
where $\hat{u}_N$ is a perturbation series
defined by (\ref{multiplescaleperturbationseries}).
\end{theorem}
{\bf Proof:} 
Let $\bar{\al }_N=(\al _0^T,\al _1^T,\cdots,\al _N^T)^T$, $\bar{\beta
 }_N=(\beta  _0^T,\beta  _1^T,
\cdots,\beta  _N^T)^T$,
$\bar{\gamma  }_N=(\gamma  _0^T,\gamma  _1^T,
\cdots,\gamma _N^T)^T\in T^*(\hat{M}_N)$,
where the sub-vectors
$\al _i,\beta _i,\gamma _i,\ 0\le i\le N$, are of the same dimension.
It suffices to prove that
\begin{equation}
<\bar{\al }_N,\hat {J}_N'({\hat{\eta}_N} )[\hat {J}_N
\bar{\beta  }_N]\bar{ \gamma }_N>+\textrm{cycle}
(\bar{\al }_N,\bar{\beta  }_N,\bar{\gamma  }_N)=0,
\end{equation}
since there is no problem on the linearity and the skew-symmetric property 
for the perturbation operator $\hat {J}_N(\hat{\eta }_N)$.

First, based on Lemma \ref{lemma:gateaxderivativeofptf}, we can calculate
the element in the $(i,j)$ position of the matrix $\hat{J}_N'({\hat{\eta}_N})
[\hat {J}_N ({\hat{\eta}_N} )\bar{\beta  }_N]$ as follows:
\begin{eqnarray} 
&&\bigl(\hat {J}_N '({\hat{\eta}_N} )[\hat {J}_N ({\hat{\eta}_N} )
\bar{\beta  }_N] \bigr)_{ij}\nonumber\\
& =& \frac 1 {(i+j-N)!}\left.\frac {\part ^{i+j-N}}{\part 
\varepsilon  ^{i+j-N}} \right|_{\varepsilon  =0} {J} '({\hat{u}_N},
\varepsilon )\Bigl[ \sum _{l=0}^N
\varepsilon  ^l (\hat {J}_N \bar{\beta  }_N)_l\Bigr]\nonumber\\
&= &\frac 1 {(i+j-N)!}\left.\frac {\part ^{i+j-N}}
{\part \varepsilon  ^{i+j-N}}\right|_{\varepsilon  =0} {J} '({\hat{u}_N},
\varepsilon  )\Bigl[ \sum _{l=0}^N\varepsilon  ^l 
 \sum_{k=N-l}^N \frac 1 {(k+l-N)!}\left.\frac {\part ^{k+l-N}
 {J}({\hat{u}_N},\varepsilon  )}{\part \varepsilon  ^{k+l-N}}
\right|_{\varepsilon  =0}\beta _k
\Bigr]\nonumber\\ &=&\frac 1 {(i+j-N)!}\left.\frac
{\part ^{i+j-N}}{\part \varepsilon  ^{i+j-N}}
\right|_{\varepsilon  =0}J '({\hat{u}_N},\varepsilon  )\Bigl[\sum_{k=0}^N
\varepsilon  ^{N-k}\bigl(\sum_{l=N-k}
^{N}\frac {\varepsilon  ^{k+l-N}}{(k+l-N)!}\left.\frac{\part ^{k+l-N}J
({\hat{u}_N} ,\varepsilon )} {\part \varepsilon  ^{k+l-N}}
\right|_{\varepsilon  =0}\bigr)\beta _k\Bigr]\nonumber\\
&=&\frac 1 {(i+j-N)!}\left.\frac {\part ^{i+j-N}}{\part 
\varepsilon  ^{i+j-N}}
\right|_{\varepsilon  =0}{J}'({\hat{u}_N},\varepsilon  )\Bigl[\sum_{k=0}^N
\varepsilon  ^{N-k}\bigl(J(\hat{u}_N ,\varepsilon )
+\textrm{o}(\varepsilon  ^k)\bigr)\beta _k\Bigr]\nonumber\\
&=&\frac 1 {(i+j-N)!}\left.\frac {\part ^{i+j-N}}
{\part \varepsilon  ^{i+j-N}}
\right|_{\varepsilon  =0} {J} '({\hat{u}_N},\varepsilon  )
\Bigl[\sum_{k=0}^N\varepsilon  ^{N-k}
J({\hat{u}_N},\varepsilon  )\beta _k\Bigr]\nonumber\\
&=&\sum_{k=0}^N\frac 1 {(i+j-N)!}\left.\frac {\part ^{i+j-N}}
{\part \varepsilon  ^{i+j-N}}
\right|_{\varepsilon  =0}\bigl(\varepsilon  ^{N-k}
{J} '({\hat{u}_N},\varepsilon  )\bigl[
J({\hat{u}_N},\varepsilon  )\beta _k\bigr]\bigr)\nonumber\\
&=&\sum_{k=2N-(i+j)}^N\frac 1 {(i+j+k-2N)!}\left.
\frac {\part ^{i+j+k-2N}}{\part \varepsilon  ^{i+j+k-2N}}
\right|_{\varepsilon  =0} {J} '({\hat{u}_N},\varepsilon  )\bigl[
J({\hat{u}_N},\varepsilon  )\beta _k\bigr],\ 0\le i,j\le N.\nonumber
\end{eqnarray}

In what follows, let us give the remaining proof for the case of
\begin{equation} \eta _i=\eta _i(y_0,y_1,t)=\eta _i(x,\varepsilon x,t),
\ 0\le i\le N. \label{twoscaleperturbationcase}\end{equation}
Suppose that the duality between cotangent vectors and tangent vectors
is given by 
\begin{equation}<\al , K> = \int_{\R^p} \al ^TK\, dx,\ x\in \R^p,\ 
\alpha \in T^*(M),\ K\in T(M). \end{equation} 
Let us consider the case of $x\in \R$ without loss of generality.
For brevity, we set 
\begin{equation}
F_{ijk}(\bar{\al }_N,\bar{\beta }_N,\bar{\gamma }_N,
\hat{\part } _x)=\Bigl(\al _i^T {J} '({\hat{u}_N},\varepsilon  )\bigl[
J({\hat{u}_N},\varepsilon  )\beta _k\bigr]\gamma _j
 +\textrm{cycle}(\al _i,\beta _k,\gamma _j)\Bigr),\ 0\le i,j,k\le N,
\end{equation}
where $\hat{\part }_x=\part _{y_0}+\varepsilon \part _{y_1}$,
owing to (\ref{twoscaleperturbationcase}), and we assume
that the original Hamiltonian operator $J(u,\varepsilon)$ 
involves the differential operator $\part _x$. Then we can have
\begin{eqnarray} &&
<\bar{\al }_N,\hat {J}_N'({\hat{\eta}_N} )[\hat {J}_N \bar{\beta  }_N]
\bar{ \gamma  }_N>+\textrm{cycle}(\bar{\al }_N,\bar{\beta  }_N,
\bar{\gamma  }_N)
\nonumber\\
&=& \int _{-\infty }^\infty \int _{-\infty }^\infty
\sum_{2N\le i+j+k\le 3N}\frac 1 {(i+j+k-2N)!}\left.
\frac {\part ^{i+j+k-2N}}{\part \varepsilon  ^{i+j+k-2N}}
\right|_{\varepsilon  =0}
F_{ijk}(\bar{\al }_N,\bar{\beta }_N,\bar{\gamma }_N,
\hat{\part } _x)\, dy_0dy_1. \nonumber \end{eqnarray} 
In order to apply the Jacobi identity of $J(u,\varepsilon)$,
we make a dependent variable transformation
\begin{equation}
y_0=p,\ y_1=q+\varepsilon p,
\end{equation} 
from which it follows that
\begin{equation}\part _p=\part _{y_0}+\varepsilon \part _{y_1},\ 
\part _q=\part _{y_1}. \end{equation} 
Now we can continue to compute that 
\begin{eqnarray}  &&
<\bar{\al }_N,\hat {J}_N'({\hat{\eta}_N} )[\hat {J}_N \bar{\beta  }_N]
\bar{ \gamma  }_N>+\textrm{cycle}(\bar{\al }_N,\bar{\beta  }_N,
\bar{\gamma  }_N) \nonumber \\ &=&
 \int _{-\infty }^\infty \int _{-\infty }^\infty
\sum_{2N\le i+j+k\le 3N}\frac 1 {(i+j+k-2N)!}\left.
\frac {\part ^{i+j+k-2N}}{\part \varepsilon  ^{i+j+k-2N}}
\right|_{\varepsilon  =0}\times \nonumber \\ && \qquad 
F_{ijk}(\bar{\al }_N,\bar{\beta }_N,\bar{\gamma }_N,{\part }_p)\left|
\rm{det}\left[\begin{array} {cc} \frac {\part y_0}{\part p}&
\frac {\part y_0}{\part q}\vspace{2mm}\\
\frac {\part y_1}{\part p}&\frac {\part y_1}{\part q}
 \end{array} \right]
\right|\, dpdq
\nonumber \\
&=& \int _{-\infty }^\infty \Bigl (
\sum_{2N\le i+j+k\le 3N}\frac 1 {(i+j+k-2N)!}\left.
\frac {\part ^{i+j+k-2N}}{\part \varepsilon  ^{i+j+k-2N}}
\right|_{\varepsilon  =0} \int _{-\infty }^\infty
F_{ijk}(\bar{\al }_N,\bar{\beta }_N,
\bar{\gamma }_N,\part _p) \, dp \Bigr) \,dq \quad \nonumber
\\ &=&
\int _{-\infty }^\infty 0 \,dq=0.\nonumber
\end{eqnarray}
In the last but one step,
we have utilized the Jacobi identity of $J(u,\varepsilon)$.

The method used here for showing the Jacobi identity can be extended 
to the other cases of the perturbations. Therefore the required result
is proved. $\vrule width 1mm height 3mm depth 0mm$

Similarly, we can show the following structure for the perturbation
symplectic operators.

\begin{theorem}
If the operator $\Theta=\Theta (\varepsilon ) :T(M)\to T^*(M)$
being analytic with respect to $\varepsilon $ is symplectic,
then the following operator
$\hat{\Theta }_N :T(\hat{M}_N)\to T^*(\hat{M}_N)$ defined by
\begin{eqnarray}&&(\textrm{per}_N\Theta )(\hat{\eta   }_N)=
\hat{\Theta }_N ({\hat{\eta}_N} )\nonumber
\\
&=&\left[ \bigl(\hat{\Theta }_N ({\hat{\eta}_N} )\bigr)_{ij}
\right]_{i,j=0,1,\cdots,N}
=\left[
\frac 1 {(N-i-j)!}\left.\frac {\part ^{N-i-j}\Theta ({\hat{u}_N},\varepsilon )}
{\part \varepsilon  ^{N-i-j}}
\right|_{\varepsilon  =0}
 \right]_{q(N+1)\times q(N+1)}\nonumber
\\
 &= &
\left[ \begin{array}{cccc}
\frac {1}{N!}\left.\frac {\part ^N \Theta ({\hat{u}_N},\varepsilon )}{\part 
\varepsilon  ^N }
\right|_{\varepsilon  =0}
&\cdots &\frac 1{1!}\left.
\frac {\part  \Theta ({\hat{u}_N} ,\varepsilon)}{\part \varepsilon  }
\right|_{\varepsilon  =0} 
 & \Theta (\hat{u}_N,\varepsilon)|_{\varepsilon=0}\vspace{1mm}\\  \vdots &
\begin{turn}{45}\vdots\end{turn}&
\begin{turn}{45}\vdots\end{turn}& \vspace{1mm} \\
\frac {1}{1!}\left.\frac {\part ^N \Theta ({\hat{u}_N},\varepsilon )}{\part 
\varepsilon  ^N }
\right|_{\varepsilon  =0} 
 &\Theta (\hat{u}_N,\varepsilon)|_{\varepsilon=0}
 & &  \vspace{2mm}\\ \Theta (\hat{u}_N,\varepsilon)|_{\varepsilon=0}& & & 0
\end{array}
\right]\label{newsym}
\end{eqnarray}is also symplectic,
where $\hat{u}_N$ is a perturbation series
defined by (\ref{multiplescaleperturbationseries}).
\end{theorem}

\subsection{Integrable Properties}
\label{integrableproperties}

In this sub-section, we study integrable properties of 
the perturbation systems defined by
(\ref{multiplescaleperturbationsystem}), which include recursion 
hereditary operators, $K$-symmetries (i.e., time independent symmetries), 
master-symmetries, Lax representations
and zero curvature representations, Hamiltonian formulations and etc.
Simultaneously we establish explicit structures for constructing 
other perturbation objects such as spectral problems,
Hamiltonian functionals, and cotangent vector fields. 

\begin{theorem}\label{thm:perturbationrecursionoperator}
Let $K=K(\varepsilon )\in T(M)$ be analytic
with respect to $\varepsilon$ and assume that $\Phi =\Phi (\varepsilon):
T(M)\to T(M)$ is a recursion operator of $u_t=K(u,\varepsilon )$.
Then the operator ${\hat{\Phi}_N}: T(\hat {M}_N)\to T(\hat {M}_N)$ 
determined by (\ref{newhere}) is a recursion operator of the perturbation 
system $\hat {\eta   }_{Nt}={\hat{K}_N} ({\hat{\eta}_N} )$ defined by 
(\ref{multiplescaleperturbationsystem}).
Therefore if $u_t=K(u,\varepsilon)$ has a hereditary recursion operator 
$\Phi (u,\varepsilon)$, then the perturbation system $\hat {\eta   }_{Nt}=
{\hat{K}_N} ({\hat{\eta}_N} )$ has a hereditary recursion operator 
${\hat{\Phi}_N}({\hat{\eta}_N} )$.
\end{theorem}
{\bf Proof:} Let $\bar{S}_N=(S_0^T,S_1^T,\cdots,S_N^T)^T\in T(\hat {M}_N)$,
where the sub-vectors $S_i,\ 0\le i\le N$, are of the same dimension.
By Lemma \ref{lemma:gateaxderivativeofptf}, we can compute that
\begin{eqnarray}
&& 
\left(\left. \frac {\part ^k\Phi ({\hat{u}_N},\varepsilon )}
{\part \varepsilon ^k}
\right|_{ \varepsilon=0}\right)'({\hat{\eta}_N} )[{\hat{K}_N} ]=\left.\frac 
{\part ^k}{\part \varepsilon ^k}\right|_{\varepsilon  =0}
\Phi '(\hat {u}_N,\varepsilon)\Bigl[\sum_{j=0}^N\varepsilon  ^j K^{(j)}\Bigr]
\nonumber\\
 &=& \left.\frac {\part ^k}{\part \varepsilon ^k}\right|_{\varepsilon  =0}
\Phi '(\hat {u}_N,\varepsilon)\bigl[K(\hat {u}_N,\varepsilon )
+\textrm{o}(\varepsilon  ^N)\bigr]=\left.
\frac {\part ^k \Phi '(\hat {u}_N,\varepsilon)[K(\hat {u}_N,\varepsilon )]}
{\part \varepsilon  ^k}\right|_{\varepsilon  =0}
,\ 0\le k\le N,\nonumber \end{eqnarray}
and 
\begin{eqnarray}
&&(K^{(i)})'({\hat{\eta}_N} )[\bar{S}_N]
=\frac 1 {i!}\left.\frac {\part ^i}{\part \varepsilon  ^i}
\right|_{\varepsilon  =0}K'({\hat{u}_N},\varepsilon )\Bigl[\sum _{k=0}^N
\varepsilon  ^kS_k\Bigr]\nonumber\\ &=& \sum_{j=0}^i\frac1 {(i-j)!}
\left.\frac {\part ^{i-j} K'({\hat{u}_N},\varepsilon )
[S_j]}{\part \varepsilon  ^{i-j}}
\right|_{\varepsilon  =0},\ 0\le i\le N.\nonumber
\end{eqnarray}
Therefore, immediately from the first equality above, we obtain
the $i$-th element of ${\hat{\Phi}_N}' ({\hat{\eta}_N} )
[{\hat{K}_N} ]\bar{S}_N$ as follows: 
\begin{equation}
\bigl( {\hat{\Phi}_N}' ({\hat{\eta}_N} )[{\hat{K}_N} ]\bar{S}_N \bigr)_i=
\sum_{j=0}^i\frac1 {(i-j)!}\left.
\frac{\part ^{i-j} (\Phi '({\hat{u}_N},\varepsilon  )
[K({\hat{u}_N} ,\varepsilon )]S_j) }
{\part \varepsilon  ^{i-j}}\right|_{\varepsilon  =0},\ 
0\le i\le N.\label{recursion1}
\end{equation}
Based on the second equality above, we can make the following computation:
\begin{eqnarray}
&&\bigl({\hat{K}_N} '({\hat{\eta}_N} )[{\hat{\Phi}_N}({\hat{\eta}_N} )
\bar {S}_N]\bigr)_i \nonumber\\
&=&\sum_{k=0}^i\frac1 {(i-k)!}\left.\frac{\part ^{i-k} }
{\part \varepsilon  ^{i-k}}\right|_{\varepsilon  =0}\bigl(
K'({\hat{u}_N},\varepsilon  )[({\hat{\Phi}_N}({\hat{\eta}_N} )\bar{S}_N)_k]
 \bigr)\nonumber\\
 &=&\sum_{k=0}^i\frac1 {(i-k)!}
\left.\frac{\part ^{i-k} }{\part \varepsilon  ^{i-k}}
\right|_{\varepsilon  =0}K'({\hat{u}_N} ,\varepsilon )
\Bigl[\sum_{j=0}^k\frac1 {(k-j)!}
\left.\frac{\part ^{k-j} \Phi ({\hat{u}_N},\varepsilon )
 }{\part \varepsilon  ^{k-j}}\right|_{\varepsilon  =0}S_j\Bigr]\nonumber\\
&=& \sum_{j=0}^i\sum_{k=j}^i\frac 1{(i-k)!(k-j)!}\left.\frac {\part ^{i-k}}{
\part \varepsilon  ^{i-k}}\right|_{\varepsilon  =0}
K'({\hat{u}_N},\varepsilon  )\Bigl[\left.
\frac{\part ^{k-j} \Phi ({\hat{u}_N} ,\varepsilon )
 }{\part \varepsilon  ^{k-j}}\right|_{\varepsilon  =0}S_j\Bigr]\nonumber\\
&=& 
\sum_{j=0}^i\sum_{k=j}^i\frac 1{(i-j)!}\left.\frac {\part ^{i-j}}{
\part \varepsilon  ^{i-j}}\right|_{\varepsilon  =0}\Bigl(\frac {\varepsilon  
^{k-j}}{(k-j)!}K'({\hat{u}_N} ,\varepsilon )\Bigl[
\left.
\frac{\part ^{k-j} \Phi ({\hat{u}_N} ,\varepsilon )
 }{\part \varepsilon  ^{k-j}}\right|_{\varepsilon  =0}S_j\Bigr]\Bigr)
\nonumber\\ &=& 
\sum_{j=0}^i\frac 1{(i-j)!}\left.\frac {\part ^{i-j}}{
\part \varepsilon  ^{i-j}}\right|_{\varepsilon  =0}
K'({\hat{u}_N},\varepsilon  )
\Bigl[\sum_{k=j}^i\frac {\varepsilon  ^{k-j}}{(k-j)!}
\left.\frac{\part ^{k-j} \Phi ({\hat{u}_N} ,\varepsilon )
 }{\part \varepsilon  ^{k-j}}\right|_{\varepsilon  =0}S_j\Bigr]\nonumber\\
 &=&\sum_{j=0}^i\frac1 {(i-j)!}
\left.\frac{\part ^{i-j} 
 }{\part \varepsilon  ^{i-j}}\right|_{\varepsilon  =0}
K'({\hat{u}_N},\varepsilon  )[
\Phi ({\hat{u}_N} )S_j+\textrm{o}(\varepsilon  ^{i-j})]\nonumber\\
&=&\sum_{j=0}^i\frac1 {(i-j)!}\left.
\frac{\part ^{i-j} \bigl(K'({\hat{u}_N},\varepsilon  )
[\Phi ({\hat{u}_N},\varepsilon  )S_j]\bigr)
 }{\part \varepsilon  ^{i-j}}\right|_{\varepsilon  =0}, \ 0\le i\le N;
\label{recursion2}
\end{eqnarray}
\begin{eqnarray}
&& \bigl({\hat{\Phi}_N}({\hat{\eta}_N} ){\hat{K}_N} '({\hat{\eta}_N} )
[\bar{S}_N]\bigr)_i\nonumber\\
&=&\sum_{k=0}^i\frac1 {(i-k)!}
\left.\frac{\part ^{i-k} \Phi ({\hat{u}_N},\varepsilon  )
 }{\part \varepsilon  ^{i-k}}\right|_{\varepsilon  =0}
\sum_{j=0}^k\frac1 {(k-j)!}
\left.\frac{\part ^{k-j} K'({\hat{u}_N},\varepsilon  )[S_j]
 }{\part \varepsilon  ^{k-j}}\right|_{\varepsilon  =0}
\nonumber\\
&=&\sum_{j=0}^i\sum_{k=j}^i\frac1 {(i-k)!(k-j)!}
\left.
\frac{\part ^{i-k} \Phi ({\hat{u}_N},\varepsilon  )
 }{\part \varepsilon  ^{i-k}}\right|_{\varepsilon  =0}
\left.
\frac{\part ^{k-j} K'({\hat{u}_N} ,\varepsilon )[S_j]
 }{\part \varepsilon  ^{k-j}}\right|_{\varepsilon  =0}\nonumber\\
&=&\sum_{j=0}^i\frac1 {(i-j)!}
\left.\frac{\part ^{i-j} \bigl(\Phi ({\hat{u}_N},\varepsilon  )
K'({\hat{u}_N} ,\varepsilon )[S_j]\bigr)
 }{\part \varepsilon  ^{i-k}}\right|_{\varepsilon  =0},\ 0\le i\le N.
\label{recursion3} \end{eqnarray}
It follows directly from the above 
three equalities above (\ref{recursion1}), (\ref{recursion2}) and
(\ref{recursion3}) that
\[\frac {\part \hat {\Phi}_N}{\part t}(\hat{\eta }_N) \bar {S}_N+
 {\hat{\Phi}_N}'({\hat{\eta}_N} )[{\hat{K}_N}(\hat{\eta }_N) ]
\bar{S}_N-{\hat{K}_N} '({\hat{\eta}_N} )
[{\hat{\Phi}_N}(\hat{\eta }_N)\bar{S}_N]
+{\hat{\Phi}_N}(\hat{\eta }_N){\hat{K}_N} '({\hat{\eta}_N} ) 
[\bar{S}_N]=0.\]
According to Definition \ref{def:recursionoperator},
this implies that the perturbation operator 
$\hat{\Phi}_N (\hat {\eta }_N)$ defined by (\ref{newhere}) 
is a recursion operator
of $\hat {\eta }_{Nt}=\hat {K}_N (\hat{\eta }_N)$. A combination 
with Theorem \ref{thm:perturbationhereditaryoperator} gives 
rise to the proof of the second required conclusion. The proof 
is finished. $\vrule width 1mm height 3mm depth 0mm$

\begin{theorem} Let $K=K(\varepsilon),S=S(\varepsilon)\in T(M)$
be analytic with respect to $\varepsilon$. For two perturbation vector fields
${\hat{K}_N},\, {\hat{S}_N}\in T(\hat{M}_N)$ defined by 
(\ref{perturbationvectorfield}), there exists the following relation:
\begin{equation}
[{\hat{K}_N} ({\hat{\eta}_N} ), {\hat{S}_N} ({\hat{\eta}_N} )]
=({\hat{K}_N} )'({\hat{\eta}_N} )[{\hat{S}_N} ({\hat{\eta}_N} )]
-({\hat{S}_N} )'({\hat{\eta}_N} )[{\hat{K}_N} ({\hat{\eta}_N} )]=
\hat {T}_N({\hat{\eta}_N} ),\label {pvfrelation}
\end{equation}
where $\hat {T}_N\in (\hat{M}_N)$ is the perturbation vector field of the 
vector field $ T(\varepsilon)=[K(\varepsilon),S(\varepsilon)]$, 
defined by (\ref{perturbationvectorfield}). Furthermore we can have
the following:

\noindent (1) 
if $\sigma =\sigma(\varepsilon)\in T(M)$
is an $n$-th order master-symmetry of the perturbed system
$u_t=K(u,\varepsilon)$,
then $\hat{\sigma}_N
\in T(\hat {M}_N)$ defined by (\ref{perturbationvectorfield})
is an $n$-th order master-symmetry of the perturbation
system $\hat {\eta}_{Nt}={\hat{K}_N} ({\hat{\eta}_N} )$ defined by
(\ref{multiplescaleperturbationsystem});

\noindent (2) the perturbation
system $\hat{\eta }_{Nt}={\hat{K}_N} ({\hat{\eta}_N} )$ 
defined by (\ref{multiplescaleperturbationsystem}) possesses
the same structure of symmetry algebras as 
the original perturbed system $u_t=K(u,\varepsilon)$.
\end{theorem}
{\bf Proof:} 
As usual, assume that 
\[ \hat {S }_i=(S^{(0)T},S^{(1)T},\cdots,S^{(i)T})^T, 
\ \hat {\eta }_i=({\eta }_0^T,{\eta }_1^T,\cdots, {\eta }_i^T)^T,\
\hat {u}_i=\sum_{k=0}^i\varepsilon ^k\eta _k,\ 0\le i\le N.\] 
By the definition of the Gateaux derivative, we first have 
\begin{eqnarray} 
&&(K(\hat {u}_i,\varepsilon ))'(\hat {\eta}_i)[\hat{S}_i]=\left.
\frac {\part }{\part \delta}\right|_{\delta =0}K(\hat {u}_i
+\delta \sum_{k=0}^i\varepsilon  ^kS^{(k)},\varepsilon )
\nonumber\\
&=& \left.\frac {\part }{\part \delta}\right|_{\delta =0}K(\hat {u}_i+\delta 
S(\hat {u}_i,\varepsilon )+\delta \textrm{o}(\varepsilon  ^i),\varepsilon )
\nonumber\\
&=& K'(\hat {u}_i,\varepsilon )[S(\hat {u}_i,\varepsilon )] 
+\textrm{o}(\varepsilon  ^i),\ 0\le i\le N.\nonumber
\end{eqnarray}
Let us apply the equality above to the following Taylor series 
\[
K(\hat {u}_i,\varepsilon )=\sum_{k=0}^i\frac{\varepsilon  ^k}{k!}
\left.\frac{\part ^k K(\hat {u}_i,\varepsilon) }{\part \varepsilon  ^k}
\right  |_{\varepsilon  =0}+\textrm{o}(\varepsilon  ^i),\ 0\le i\le N,\]
and then we arrive at 
\[K'(\hat{u}_i,\varepsilon)[S(\hat{u}_i,\varepsilon)]=
\sum_{k=0}^i\frac {\varepsilon ^k}{k!} \Bigl(\left .\frac 
{\part ^kK(\hat{u}_i,\varepsilon)}{\part \varepsilon  ^k}
\right|_{\varepsilon=0}\Bigr)' (\hat {\eta   }_i)[\hat{S}_i]
+\textrm{o}(\varepsilon ^i),\ 0\le i\le N. \]
Taking the $i$-th derivative with respect to $\varepsilon $ leads to 
\begin{equation} 
\bigr(K'(u,\varepsilon)[S(u,\varepsilon )]\bigl)^{(i)}(\hat {\eta   }_i)
=\Bigr(\bigr(K(u,\varepsilon )
\bigl)^{(i)}
\Bigl)'(\hat {\eta   }_i)[\hat {S }_i],\ 0\le i\le N.
\label{pvfcomponentrelation}
\end{equation}

Now it follows from (\ref{pvfcomponentrelation}) that 
for the $i$-th element of $\hat{T}_N$ we have 
\begin{eqnarray} &&(T(u,\varepsilon ))^{(i)}=
\bigl(K'(u,\varepsilon )[S(u,\varepsilon )]\bigr)^{(i)}(\hat {\eta   }_i)-
\bigl(S'(u,\varepsilon )[K(u,\varepsilon )]\bigr)^{(i)}
(\hat {\eta   }_i)\nonumber\\ &=&
\bigl(\bigl(K(u,\varepsilon )\bigr)^{(i)}\bigr)'(\hat {\eta   }_i )
[\hat {S }_i]-\bigl(\bigl(S(u,\varepsilon )\bigr)
^{(i)}\bigr)'(\hat {\eta   }_i )[\hat {K}_i]\nonumber\\
 &=&
\bigl({\hat{K}_N} '({\hat{\eta}_N} )[\hat {S}_N]\bigr)_i-\bigl({\hat{S}_N}
  '({\hat{\eta}_N} )[\hat {K}_N]\bigr)_i
,\ 0\le i\le N.\nonumber
\end{eqnarray} 
This shows that (\ref{pvfrelation}) holds. All other results
is a direct consequence of (\ref{pvfrelation}). 
The proof is completed. $\vrule width 1mm height 3mm depth 0mm$

The relation (\ref{pvfrelation}) implies that the perturbation series 
(\ref{multiplescaleperturbationseries}) 
keeps the Lie product of vector fields invariant.
In particular, the second component of (\ref{pvfrelation}) yields 
the Lie product property (\ref{lieproductoflinearizeoperators}) 
of linearized operators. In what follows, we will go on to
consider Lax representations and zero curvature representations
for the perturbation system defined by
(\ref{multiplescaleperturbationsystem}).
In our formulation below, we will adopt the following notation for
the perturbation of a spectral parameter $\lambda $:
\begin{equation} \hat {\lambda }_N=\sum_{i=0}^N \varepsilon
^i \mu _i ,\ \hat {\mu }_N =(\mu _0,\mu _1,\cdots ,\mu _N)^T ,
\label{perturbationofspectralparameter}
\end{equation}
which is quite similar to the notation for the perturbation of the potential
$u$. Here $\mu _i,\ 0\le i\le N,$ will be taken as the spectral parameters
appearing in the perturbation spectral problems. A customary symbol 
$\bigtriangledown _x\lambda ,\ x\in \R ^p,$ will still be used to 
denote the gradient of the spectral parameter $\lambda $ with respect to $x$.

\begin{theorem} \label{pelaxrepresentation}
Let $K=K(\varepsilon)\in T(M)$ be analytic with respect to
$\varepsilon$. Assume that 
the system $u_t=K(u,\varepsilon)$ has an isospectral Lax representation
\begin{equation}\left\{ \begin{array} {l}L(u,\varepsilon)\phi =\lambda \phi, 
\vspace{2mm}\\ \phi _t=A(u,\varepsilon)\phi, 
\end{array} \right.\ (\bigtriangledown _x\lambda =0,\ x\in \R ^p),\ 
\textrm{i.e.,}\ 
(L(u,\varepsilon))_t=[A(u,\varepsilon),L(u,\varepsilon)], 
\label{laxrepresentationofu_t=K(uvarepsilon)}\end{equation} 
where $L$ and $A$ are two $s\times s$ matrix differential operators
being analytic with respect to $u$ and $\varepsilon$.
Define the perturbation spectral operator $\hat {L}_{N}$ and the perturbation 
Lax operator $\hat {A}_{N}$ by 
\begin{eqnarray}
&&(\textrm{per}_NB)(\hat {\eta   }_N)=
\hat {B}_{N}({\hat{\eta}_N} )\nonumber
\\ &=&\left[\bigl(\hat {B}_{N}({\hat{\eta}_N} )\bigr)_{ij}
\right]_{i,j=0,1,\cdots,N}=\left[\left.
\frac 1 {(i-j)!}\frac {\part ^{i-j}B ({\hat{u}_N} ,\varepsilon)}
{\part \varepsilon  ^{i-j}} \right|_{\varepsilon  =0}
 \right]_{s(N+1)\times s(N+1)}\nonumber
\\ &= &
\left[ \begin{array}{cccc}B (\hat{u}_N,\varepsilon)|_{\varepsilon=0}
& & &0\\ \left.
\frac {1}{1!}\frac {\part B ({\hat{u}_N} ,\varepsilon)}{\part \varepsilon  }
\right|_{\varepsilon  =0}&B (\hat{u}_N,\varepsilon)|_{\varepsilon=0} & 
& \\ \vdots & \ddots & \ddots & \\
 \left. \frac {1}{N!}\frac {\part ^N B ({\hat{u}_N},\varepsilon )}
{\part \varepsilon  ^N} \right|_{\varepsilon  =0}&\cdots &\left.
\frac {1}{1!}\frac {\part B ({\hat{u}_N},\varepsilon )}{\part \varepsilon  }
\right|_{\varepsilon  =0}&B (\hat{u}_N,\varepsilon)|_{\varepsilon=0}
\end{array}\right], \ B=L,A,\qquad\ \ 
\end{eqnarray}
where $\hat{u}_N$ is given by (\ref{multiplescaleperturbationseries}).
Then under the condition for the spectral operator $L$ that
\begin{equation}
 \textrm{if}\  L'(\hat{u}_N)[S(\hat{u}_N)]=\textrm{o}(\varepsilon ^N),\
 S\in T(M),\  
\textrm{then}\  S(\hat{u}_N)=\textrm{o}(\varepsilon ^N),
\label{injectiveconditionuptovarepsilon^NforL}\end{equation} 
the $N$-th order perturbation system $\hat {\eta   }_{Nt}=
{\hat{K}_N} ({\hat{\eta}_N} )$ defined by 
(\ref{multiplescaleperturbationsystem}) has the following isospectral 
Lax representation
\begin{equation} (\hat {L}_N({\hat{\eta}_N} ))_{t}
=[\hat {A}_{N}({\hat{\eta}_N} ),\hat {L}_{N}({\hat{\eta}_N} )],
\label{perturbationlaxequation} \end{equation} 
which is the compatibility condition of 
the following perturbation spectral problem
\begin{equation}
\left\{ \begin{array} {l}\hat {L}_N({\hat{\eta}_N} )\hat{\phi}_N =\lambda
\hat{\phi}_N, \vspace{2mm}\\
\hat{\phi}_{Nt}=\hat {A}_{N}({\hat{\eta}_N} )\hat{\phi}_N, \end{array} \right.
(\bigtriangledown _{y_0} \lambda =\
\bigtriangledown _{y_1} \lambda =\cdots =
\bigtriangledown _{y_r} \lambda =0),
\label{perturbationlaxspectralproblem1}
\end{equation}
or the following perturbation spectral problem
\begin{equation}
\left\{ \begin{array} {l}\hat {L}_N({\hat{\eta}_N} )\hat{\phi}_N =\Lambda
\hat{\phi}_N, \vspace{2mm}\\ \hat{\phi}_{Nt}=\hat {A}_{N}
({\hat{\eta}_N} )\hat{\phi}_N, \end{array} \right. 
\label{perturbationlaxspectralproblem2}
\end{equation}
where the matrix $\Lambda $ reads as
\begin{equation}\Lambda =\left [ \begin{array} {cccc} \mu  _0I_s & & &
\vspace{2mm}\\
\mu  _1 I_s&\mu  _0I_s & & \vspace{2mm}\\
\vdots & \ddots & \ddots & \vspace{2mm}\\
\mu  _N I_s&\cdots &\mu  _1I_s & \mu  _0I_s 
\end{array} \right],\ I_s=\textrm{diag}({\underbrace
{1,1,\cdots,1} _{s} }), \end{equation} 
with the spectral parameter $\mu _i,\ 0\le i\le N$, satisfying 
\begin{equation}
\sum_{k+l=i}\bigtriangledown _{y_k}\mu _l=0,\ 0\le i\le N.
 \label{generalconditionsofspectralparameters}
\end{equation} 
\end{theorem}
{\bf Proof:} 
We first observe that the perturbed system
\begin{equation}\hat {u}_{Nt}=K(\hat{u}_N,\varepsilon)+\textrm{o}
(\varepsilon  ^N),\label{pertuebationsystemuptovarepsilon^N}\end{equation}
which engenders precisely the perturbation system
$\hat{\eta }_{Nt}= \hat {K}_N(\hat {\eta }_N)$ defined by
(\ref{multiplescaleperturbationsystem}).
Noting that $L(u,\varepsilon),\,A(u,\varepsilon)$ are analytic 
with respect to $u$ and $\varepsilon$, it follows from 
(\ref{laxrepresentationofu_t=K(uvarepsilon)}) that 
(\ref{pertuebationsystemuptovarepsilon^N}) is equivalent to the following
\begin{equation}
\left.\frac{\part ^{k}}{\part \varepsilon  ^{k}}\right|_{\varepsilon  =0}
\Bigl((L(\hat{u}_N,\varepsilon))_t-[A(\hat{u}_N,\varepsilon),
L(\hat{u}_N,\varepsilon)]\Bigr)=0,\ 0\le k\le N,
\label{laxrep0toN}
\end{equation}
by use of (\ref{injectiveconditionuptovarepsilon^NforL}).

What we want to prove next is that (\ref{laxrep0toN}) is equivalent to 
(\ref{perturbationlaxequation}). 
Let us compute the elements of the differential operator 
matrix $[\hat {A}_N(\hat{\eta }_N), \hat{L}_N(\hat{\eta }_N)]$.
It is obvious that $[\hat {A}_N(\hat{\eta }_N), \hat{L}_N(\hat{\eta }_N)]$ 
is lower triangular, that is to say,
\[ ([\hat {A}_N(\hat{\eta }_N), \hat{L}_N(\hat{\eta }_N)])_{ij}=0,
\ 0\le i<j\le N.\]
For the other part of 
$[\hat {A}_N(\hat{\eta }_N), \hat{L}_N(\hat{\eta }_N)]$, we can compute that
\begin{eqnarray}
(\hat {A}_N(\hat{\eta }_N) \hat{L}_N(\hat{\eta }_N))_{ij}
&=&\sum_{k=j}^i\frac{1}{(i-k)!}\left.\frac{
\part ^{i-k}A(\hat{u}_N,\varepsilon)}{\part \varepsilon  
^{i-k}}\right|_{\varepsilon  =0}
\frac{1}{(k-j)!}\left.\frac{
\part ^{k-j}L(\hat{u}_N,\varepsilon)}
{\part \varepsilon  ^{k-j}}\right|_{\varepsilon  =0}\nonumber\\
&=& \frac 1{(i-j)!}\sum_{k=j}^i{ i-j \choose i-k}
\left.\frac{\part ^{i-k}A(\hat{u}_N,\varepsilon)}{\part
 \varepsilon  ^{i-k}}\right|_{\varepsilon  =0}
\left.\frac{\part ^{k-j}L(\hat{u}_N,\varepsilon)}
{\part \varepsilon  ^{k-j}}\right|_{\varepsilon  =0}\nonumber
\\
&=& \frac 1{(i-j)!}\left.\frac{
\part ^{i-j}A(\hat{u}_N,\varepsilon)L(\hat{u}_N,\varepsilon)}
{\part \varepsilon  ^{i-j}}\right|_{\varepsilon  =0},\ 
0\le j\le i\le N, \nonumber
\end{eqnarray}
where the $i\choose j$ are the binomial coefficients.
In the same way, we can obtain
\[(\hat {L}_N(\hat{\eta }_N) \hat{A}_N(\hat{\eta }_N))_{ij}=
\frac 1{(i-j)!}\left.\frac{
\part ^{i-j}L(\hat{u}_N,\varepsilon)A(\hat{u}_N,\varepsilon)}
{\part \varepsilon  ^{i-j}}\right|_{\varepsilon  =0},
\ 0\le j\le i\le N.\]
Therefore we arrive at
\[([\hat {A}_N(\hat{\eta }_N), \hat{L}_N(\hat{\eta }_N)])_{ij}=
\frac 1{(i-j)!}\left.\frac{
\part ^{i-j}[A(\hat{u}_N,\varepsilon),L(\hat{u}_N,\varepsilon)]}
{\part \varepsilon  ^{i-j}}\right|_{\varepsilon  =0},
\ 0\le j\le i\le  N.\]
Now it is easy to find that (\ref{laxrep0toN}) is equivalent to
(\ref{perturbationlaxequation}). Therefore the perturbation system
defined by (\ref{multiplescaleperturbationsystem}) has the Lax
representation (\ref{perturbationlaxequation}). 

Let us now turn to the perturbation spectral problems
(\ref{perturbationlaxspectralproblem1}) and
(\ref{perturbationlaxspectralproblem2}).
Obviously, the compatibility condition of the perturbation spectral problem
(\ref{perturbationlaxspectralproblem1}) is the Lax equation
(\ref{perturbationlaxequation}),
since the spectral parameter $\lambda $ doesn't vary whatever the spatial
variables change. Therefore let us consider 
the compatibility condition of the perturbation spectral problem
(\ref{perturbationlaxspectralproblem2}).
First, we want to prove that
\begin{equation}\Lambda \hat {A}_N(\hat{\eta }_N)=
 \hat {A}_N(\hat{\eta }_N)\Lambda ,
\label{commutativepropertyofLambdahatA{N}}\end{equation} 
if the spectral parameters $\mu _i,\ 0\le i\le N,$ satisfy
(\ref{generalconditionsofspectralparameters}).
Notice that the condition (\ref{generalconditionsofspectralparameters})
on the spectral parameters $\mu _i,\ 0\le i\le N$, is required by
\[\hat {\bigtriangledown }_x\hat {\lambda }_N=\textrm{o}(\varepsilon  ^N),
\ \hat {\bigtriangledown }_x=\sum_{i=0}^r\varepsilon ^i \bigtriangledown
_{y_i},\ \hat {\lambda }_N= \sum_{i=0}^N\varepsilon ^i\mu  _i, \]
which is a perturbation version of $\bigtriangledown _x\lambda =0$.
Therefore we have 
\[A(\hat {u}_N,\varepsilon, \hat {\bigtriangledown  }_x)
\hat {\lambda }_N=\hat {\lambda }_N 
A(\hat {u}_N,\varepsilon, \hat {\bigtriangledown }_x)
+\textrm{o}(\varepsilon  ^N).\]
This guarantees that
\[ \frac 1{(i-j)!} 
\left.\frac{\part ^{i-j}}{\part \varepsilon  ^{i-j}}\right|_{\varepsilon  =0}
\Bigl(A(\hat {u}_N,\varepsilon, \hat {\bigtriangledown}_x)
\hat {\lambda }_N\Bigr)
=\frac 1{(i-j)!} 
\left.\frac{\part ^{i-j}}{\part \varepsilon  ^{i-j}}\right|_{\varepsilon  =0}
\Bigl(\hat {\lambda }_N A(\hat {u}_N,\varepsilon,
\hat {\bigtriangledown }_x)\Bigr),\ 0\le j\le i\le N,
\]
which exactly means that 
the equality (\ref{commutativepropertyofLambdahatA{N}}) holds.
Now we can compute from (\ref{perturbationlaxspectralproblem2}) that
\[ (\hat {L}_N(\hat{\eta}_N))_{t}\hat{\phi}_N+\hat {L}_N(\hat{\eta}_N)
\hat {A}_N(\hat{\eta}_N)\hat {\phi}_N=
\Lambda \hat {A}_N(\hat{\eta}_N)\hat {\phi }_N=\hat {A}_N(\hat{\eta}_N)\Lambda
\hat {\phi }_N=\hat {A}_N(\hat{\eta}_N)\hat {L}_N(\hat{\eta}_N)\hat {\phi}_N.
\] 
It follows that the compatibility condition of the perturbation spectral
problem (\ref{perturbationlaxspectralproblem2}) is also
the Lax equation (\ref{perturbationlaxequation}). The proof is completed.
$\vrule width 1mm height 3mm depth 0mm$

The perturbation spectral operator $\hat {L}_N$ is very similar
to the perturbation recursion operator $\hat {\Phi }_N$, in spite of
different orders of matrices. Actually,
we may take any recursion operator $\Phi $ as a spectral operator and 
the system $u_t=K(u)$ can have a Lax representation $\Phi _t=[\Phi ,K']$.
This Lax representation is usually non-local, because
most recursion operators are intego-differential.
We also remark that two perturbation spectral problems above are
represented for the same perturbation system defined by
(\ref{multiplescaleperturbationsystem}),
which involve different conditions on the spectral parameters.
For the case of
\begin{equation} \hat {u}_N=\sum_{i=0}^N\varepsilon ^i\eta _i(x,y,t)=
\sum_{i=0}^N\varepsilon ^i\eta _i(x,\varepsilon x,t), \ x\in \R ,
\label{specificcaseofperturbation} \end{equation}
the condition (\ref{generalconditionsofspectralparameters}) can be
reduced to 
\begin{equation} \mu  _{0x}=0,\ \mu  _{ix}+\mu  _{i-1,y}=0, \ 1\le i\le N.
 \label{conditionsofspectralparameters}
\end{equation}
In the following theorem,
a similar result is shown for zero curvature 
representations of the perturbation systems defined by 
(\ref{multiplescaleperturbationsystem}).

\begin{theorem} 
\label{thm:zcrepofpe}
Let $K=K(\varepsilon)\in T(M)$ be analytic with respect to $\varepsilon $.
Assume that the initial system $u_t=K(u,\varepsilon)$ has 
an isospectral zero curvature representation
\begin{equation} \left \{\begin{array} {l} \phi _x=
U(u,\lambda ,\varepsilon)\phi, \vspace{2mm}
\\ \phi _t=V(u,\lambda ,\varepsilon)\phi ,\end{array} \right.
\ (\lambda _x=0,\ x\in \R),
\end{equation}
\begin{equation} \textrm{i.e.,}\ 
(U(u,\lambda ,\varepsilon))_t-(V(u,\lambda ,\varepsilon))_x+[U(u,
\lambda ,\varepsilon),V(u,\lambda ,\varepsilon)]=0, \end{equation}
where $U$ and $V$ are two $s\times s$ matrix differential 
(sometimes multiplication) operators being analytic with respect to 
$u$, $\lambda $ and $\varepsilon$. Define two perturbation matrix 
differential operators $\hat {U}_{N}$ and $\hat {V}_{N}$ by
\begin{eqnarray}
&&(\textrm{per}_N W  )(\hat {\eta   }_N)=\hat { W  }_{N}({\hat{\eta}_N},
\hat {\mu }_N )= \hat { W  }_{N}({\hat{\eta}_N} )\nonumber
\\
&=&\left[\bigl(\hat { W  }_{N}({\hat{\eta}_N} )\bigr)_{ij}
\right]_{i,j=0,1,\cdots,N}=\left[\left.
\frac 1 {(i-j)!}\frac {\part ^{i-j} W   ({\hat{u}_N},
\hat{\lambda }_N ),\varepsilon}{\part \varepsilon  ^{i-j}}
\right|_{\varepsilon  =0}
 \right]_{s(N+1)\times s(N+1)}\nonumber
\\  &= &
\left[ \begin{array}{cccc}
 W   (\hat{\eta}_N, \hat {\lambda }_N,\varepsilon )
\Bigl.\Bigr|_{\varepsilon=0} & & &0\vspace{1mm}\\ \left.
\frac {1}{1!}\frac {\part W ({\hat{u}_N}, \hat {\lambda }_N,\varepsilon  )}
{\part \varepsilon  } \right|_{\varepsilon  =0}&
 W   (\hat{\eta}_N, \hat {\lambda }_N ,\varepsilon)
 \Bigl.\Bigr|_{\varepsilon=0} & &\vspace{1mm} \\
\vdots & \ddots & \ddots & \vspace{1mm}\\
 \left. \frac {1}{N!}\frac {\part ^N  W
 ({\hat{u}_N}, \hat {\lambda }_N,\varepsilon  )} {\part \varepsilon  ^N}
\right|_{\varepsilon  =0}&\cdots &\left.
\frac {1}{1!}\frac {\part  W   ({\hat{u}_N},\hat {\lambda }_N ,\varepsilon )}
{\part \varepsilon  } \right|_{\varepsilon  =0}&
 W   (\hat{\eta}_N,\hat {\lambda }_N,\varepsilon )
\Bigl.\Bigr|_{\varepsilon=0} \end{array}
\right], \qquad\quad
\end{eqnarray}
where $W=U,V$, and $\hat{u}_N$ and $\hat{\lambda }_N$ are given by 
(\ref{specificcaseofperturbation}) and 
(\ref{perturbationofspectralparameter}). 
Then under the condition for the spectral operator $U$ that 
\begin{equation}
 \textrm{if}\ U'(\hat{u}_N)[S(\hat{u}_N)]=\textrm{o}(\varepsilon ^N),\
S\in T(M),\ \textrm{then}\  S(\hat{u}_N)=\textrm{o}(\varepsilon ^N),
\label{injectiveconditionuptovarepsilonNforU}
\end{equation}
the $N$-th order perturbation system $\hat {\eta   }_{Nt}={\hat{K}_N} 
({\hat{\eta}_N} )$ defined by (\ref{multiplescaleperturbationsystem})
has the following isospectral zero curvature representation
\begin{equation}
\left\{\begin{array} {l} \displaystyle
{\sum_{i=0}^r} \Pi ^i \hat {\phi }_{Ny_i}
=\hat {U}_N(\hat {\eta }_N,\hat{\mu}_N ) \hat {\phi }_N,
\vspace{2mm}\\ \hat {\phi }_{Nt}=\hat {V}_N(\hat {\eta }_N,\hat{\mu}_N )
\hat {\phi }_N,\end{array}  \right.
\label{perturbationzcspectralproblem}\end{equation}
\begin{equation}\textrm{i.e.,} \ (\hat {U}_N({\hat{\eta}_N} ))_{t}
-\sum_{i=0}^r \Pi ^i (\hat {V}_{N}(\hat{\eta }_N))_{y_i}
+[\hat {U}_{N}({\hat{\eta}_N} ),
\hat {V}_{N}({\hat{\eta}_N} )]=0,
\label{perturbationzcrep} \end{equation} 
where the matrix $\Pi $ is defined by
\begin{equation}
 \Pi =\left [\begin{array} {cc} 0&0\vspace{2mm}\\
I_{sN} &0 \end{array}  \right ]_{s(N+1)\times s(N+1)},\ 
I_{sN}=\textrm{diag}(\underbrace{I_s,\cdots ,I_s}_{N})
=\textrm{diag}(\underbrace{1,\cdots ,1}_{sN}),\end{equation}
and the spectral parameters $\mu_i,\ 0\le i\le N,$
satisfy
 \begin{equation} \sum_{k+l=i}\part _{y_k}\mu _l=0,\ 0\le i\le N.
\label{generalconditionsofspectralparameters2}  \end{equation}
\end{theorem}
{\bf Proof:}
Note that by use of (\ref{injectiveconditionuptovarepsilonNforU}),
the zero curvature equation 
\[ ((U({u},\lambda ,\varepsilon))_t-(V({u},\lambda ,\varepsilon))_x
+[U({u},{\lambda },\varepsilon),V( {u},\lambda ,\varepsilon)]=0\]
for the system $u_t=K(u,\varepsilon)$ yields an equivalent representation
\begin{eqnarray} &&   (U(\hat {u}_N,\hat {\lambda }_N,\varepsilon))_t
-\sum_{i=0}^r \varepsilon ^i (V(\hat {u}_N,\hat {\lambda }_N,
\varepsilon))_{y_i} +\left[U(\hat {u}_N,\hat {\lambda }_N,\varepsilon),
V(\hat {u}_N,\hat {\lambda }_N,\varepsilon)\right] \nonumber \\
&\equiv  &
U'(\hat {u}_N)[K(\hat{u}_N,\varepsilon)]
- \sum_{i=0}^r \varepsilon ^i (V(\hat {u}_N,\hat {\lambda }_N,
\varepsilon))_{y_i}
+[U(\hat {u}_N,\hat {\lambda }_N,\varepsilon),
V(\hat {u}_N,\hat {\lambda }_N,\varepsilon)] \nonumber \\
&\equiv & \textrm{o}(\varepsilon ^N) \ \pmod{\varepsilon  ^N}
\label{basicpezcrep} \end{eqnarray}
for the perturbation system $\hat{\eta }_{Nt}=\hat {K}_N(\hat {\eta }_N)$.
In order to recover $\hat {u}_{Nt}=K(\hat {u}_N,\varepsilon)+\textrm{o}
(\varepsilon^N) $ from (\ref{basicpezcrep}), we need to keep 
the spectral property $\lambda _x=\part _x\lambda =0$ under the perturbation
up to a precision $\textrm{o}(\varepsilon ^N)$.
This requires 
\[ \hat {\part }_x \hat {\lambda }_N=\textrm{o}(\varepsilon ^N),\
\hat{\part }_x= \sum _{i=0}^r \varepsilon ^i\part _{y_i}, \hat {\lambda }_N=
\sum_{i=0}^N \varepsilon ^i \mu _i, \]
which generates (\ref{generalconditionsofspectralparameters2}).
Similar to the proof of Theorem \ref{pelaxrepresentation},
differentiating the above equation (\ref{basicpezcrep}) with respect to
$\varepsilon$ up to $N$ times leads to 
the zero curvature equation (\ref{perturbationzcrep}), and 
conversely, we have (\ref{basicpezcrep}) if (\ref{perturbationzcrep}) holds.
Therefore the perturbation system
$\hat{\eta }_{Nt}=\hat {K}_N(\hat {\eta }_N)$ has an isospectral zero 
curvature representation (\ref{perturbationzcrep}).
                                    
The other thing that we need to prove is that the zero curvature equation
(\ref{perturbationzcrep}) is exactly the compatibility condition
of the perturbation spectral problem (\ref{perturbationzcspectralproblem}).
From the first system of (\ref{perturbationzcspectralproblem}), we have 
\[ \sum_{i=0}^r\Pi ^i\hat{\phi}_{Ny_it}=\hat {U}_{Nt}\hat{\phi}_N
+\hat {U}_N\hat{\phi}_{Nt}.
 \]
From the second system of (\ref{perturbationzcspectralproblem}), we obtain 
\[\hat{\phi}_{Nty_i}=\hat {V}_{Ny_i}\hat{\phi}_N+\hat {V}_N
\hat{\phi}_{Ny_i}, \  0\le i\le r. \]
A combination of the above equalities yields 
\begin{equation}
\sum_{i=0}^r\Pi ^i(\hat{V}_{Ny_i}\hat{\phi}_N+\hat {V}_N\hat{\phi}_{Ny_i})
=\hat {U}_{Nt}\hat{\phi}_N+\hat {U}_N\hat {V}_N
\hat{\phi}_N. \label{hatVNhatUN}\end{equation}
On the other hand, we have
\begin{equation}
\sum_{i=0}^N \Pi ^i \hat {V}_N \hat{\phi}_{Ny_i}
=\sum_{i=0}^N \hat {V}_N \Pi ^i\hat{\phi} _{Ny_i}
=\hat {V}_N\hat {U}_N\hat{\phi}_N,\label{HatVNtimesspatialpart}
\end{equation}
by using $\Pi \hat {V}_N=\hat {V}_N\Pi $ and the first system of 
(\ref{perturbationzcspectralproblem}).
It follows from (\ref{hatVNhatUN}) and (\ref{HatVNtimesspatialpart}) that 
the zero curvature equation (\ref{perturbationzcrep}) is 
the compatibility condition of the perturbation spectral problem
(\ref{perturbationzcspectralproblem}). The proof is completed.
$\vrule width 1mm height 3mm depth 0mm$

If we consider the specific case of the perturbation defined by
(\ref{specificcaseofperturbation}), then the perturbation
spectral problem and the perturbation zero curvature
equation, defined by (\ref{perturbationzcspectralproblem})
and (\ref{perturbationzcrep}), will be simplified to
\begin{equation}
\left\{\begin{array} {l} \hat {\phi }_{Nx}+\Pi \hat {\phi }_{Ny}=
\hat {U}_N(\hat {\eta }_N,\hat{\mu}_N ) \hat {\phi }_N,
\vspace{2mm}\\ \hat {\phi }_{Nt}=\hat {V}_N(\hat {\eta }_N,\hat{\mu}_N )
\hat {\phi }_N,\end{array}  \right.
\label{specificperturbationzcspectralproblem}\end{equation}
and 
\begin{equation} (\hat {U}_N({\hat{\eta}_N} ))_{t}
-(\hat {V}_N({\hat{\eta}_N} ))_x
-\Pi (\hat {V}_{N}(\hat{\eta }_N))_y +[\hat {U}_{N}({\hat{\eta}_N} ),
\hat {V}_{N}({\hat{\eta}_N} )]=0,
\label{specificperturbationzcrep} \end{equation} 
respectively. The involved spectral parameters $\mu _i,\ 0\le i\le N,$
need to satisfy a reduction (\ref{conditionsofspectralparameters})
of the general condition (\ref{generalconditionsofspectralparameters2}).

\begin{theorem} \label{thm:perturbationbi-Hamiltonianformulation}
Let $K=K(\varepsilon)\in T(M)$ be analytic with respect to
$\varepsilon$.
Assume that the initial system $u_t=K(u,\varepsilon)$ 
possesses a Hamiltonian formulation
\[ u_t=K(u,\varepsilon)=J(u,\varepsilon)\frac {\delta \tilde{H}}
{\delta u}(u,\varepsilon), 
\]
where $J:T^*(M)\to T(M)$ is a Hamiltonian operator and 
$\tilde{H}\in C^\infty (M)$ is a Hamiltonian functional.
Then the perturbation system $\hat {\eta} _{Nt}={\hat{K}_N} 
({\hat{\eta}_N} )$ defined by (\ref{multiplescaleperturbationsystem}) 
also possesses a Hamiltonian formulation 
\begin{equation}
\hat {\eta} _{Nt}={\hat{K}_N} ({\hat{\eta}_N} )=
{\hat{J}_N} ({\hat{\eta}_N} )\frac 
{\delta ( \textrm{per}_N \tilde{H}) }{\delta \hat{\eta}_N }
({\hat{\eta}_N} ),
\end{equation} 
where the Hamiltonian operator ${\hat{J}_N} ({\hat{\eta}_N} )$ 
is determined by (\ref{msperHamiltonianoperator}) 
and the Hamiltonian functional $\textrm{per}_N \tilde{H}
=\Hat {\tilde{{H}}}_N\in C^\infty (\hat{M}_N)$ is defined by
\begin{equation}
(\textrm{per}_N\tilde{H})(\hat{\eta   }_N)=
\Hat {\tilde{{H}}}_N({\hat{\eta}_N} )= 
\frac 1{N!}\left. \frac {\part ^N \tilde{H}({\hat{u}_N} ,\varepsilon)}
{\part \varepsilon  ^N}\right |_{\varepsilon  =0}.\label{constantsformular}
\end{equation}
The corresponding Poisson bracket has the property
\begin{equation}\{ \textrm{per}_N\tilde{H}_{1},\textrm{per}_N \tilde{H}_{2}\}
_{\hat {J}_{N}}=\textrm{per}_N 
\{\tilde{H}_1,\tilde{H}_2\}
_J,\ \tilde{H}_1,\tilde{H}_2\in C^\infty (M).
\label{Poissonpro}\end{equation} 
Moreover the perturbation systems 
$\hat {\eta}_{Nt}={\hat{K}_N} ({\hat{\eta}_N} )$
defined by (\ref{multiplescaleperturbationsystem})
possesses a multi-Hamiltonian formulation
\[
\hat {\eta} _{Nt}={\hat{K}_N} ({\hat{\eta}_N} )=
\hat{J}_{1N}({\hat{\eta}_N} )\frac{\delta (\textrm{per}_N\tilde{H}_1)}
{\delta {\hat{\eta}_N} }({\hat{\eta}_N} )=
\cdots=
\hat{J}_{mN}({\hat{\eta}_N} )\frac{\delta (\textrm{per}_N\tilde{H}_{m})}
{\delta {\hat{\eta}_N} }({\hat{\eta}_N} ),
\]
if $u_t=K(u,\varepsilon)$ possesses an analogous
 multi-Hamiltonian formulation
\[u_t=K(u,\varepsilon )=J_1(u,\varepsilon )\frac{\delta \tilde{H}_1
}{\delta u}(u ,\varepsilon )=
\cdots=
J_m(u,\varepsilon )\frac{\delta \tilde{{H}}_m}{\delta u}(u ,\varepsilon ).\]
\end{theorem}
{\bf Proof:} Assume that $\gamma (\varepsilon) 
=\frac {\delta \tilde{H}}{\delta u}(\varepsilon )\in T^*(M)$. 
Let us observe that
\begin{eqnarray}
\eta   _{it}&=& \frac1 {i!}\left.\frac {\part ^i(J({\hat{u}_N},\varepsilon )
\gamma ({\hat{u}_N},\varepsilon  ))}{\part \varepsilon  ^i}
\right|_{\varepsilon  =0}\nonumber \\
&=& \sum_{j=0}^i\frac 1{j!(i-j)!}
\left.\frac {\part ^{i-j} J({\hat{u}_N},\varepsilon  )}
{\part \varepsilon  ^{i-j}}
\right|_{\varepsilon  =0}\left.\frac {\part ^j \gamma 
({\hat{u}_N},\varepsilon  )}{\part \varepsilon  ^j}
\right|_{\varepsilon  =0},\ 0\le i\le N. \nonumber 
\end{eqnarray}
Thus, noting the structure of $\hat {J}_N$, we can represent 
the perturbation system as follows 
\begin{equation}
\hat {\eta}_{Nt}={\hat{K}_N} ({\hat{\eta}_N} )={\hat{J}_N}
 (\hat {\eta}_N)\hat {\gamma }_N(\hat {\eta}_N),
\label{newhamiltonianstru}
\end{equation}
where the cotangent vector field 
$\hat {\gamma }_N\in T^*(\hat {M}_N)$ reads as
\begin{eqnarray}  && \hat {\gamma }_N(\hat {\eta}_N)=
\Bigl( \frac 1 {N!}\left.\frac 
{\part ^N\gamma ^T({\hat{u}_N} ,\varepsilon )}
{\part \varepsilon  ^N}\right|_{\varepsilon  =0}
,\frac 1 {(N-1)!}\left.\frac 
{\part ^{N-1}\gamma ^T({\hat{u}_N},\varepsilon  )}{\part \varepsilon  ^{N-1}}
\right|_{\varepsilon  =0}
, \nonumber \\ &&\quad \qquad \qquad 
\cdots, \frac 1 {1!}\left.\frac 
{\part \gamma ^T({\hat{u}_N},\varepsilon  )}
{\part \varepsilon  }\right|_{\varepsilon  =0}
, \gamma ^T(\hat{u}_N,\varepsilon )|_{\varepsilon =0}\Bigr)^T.\label{pcvf}
\end{eqnarray} 
Let us check whether this cotangent vector field
 $\hat {\gamma }_N$ is a gradient field. If it is gradient,
the corresponding potential functional has to be the following
\begin{eqnarray} &&\tilde{H}_N({\hat{\eta}_N} )
=\int _0^1<\hat {\gamma }_N(\lambda {\hat{\eta}_N} ), {\hat{\eta}_N} >\,
d\lambda \nonumber\\&=&
\int_0^1\sum_{i=0}^N\frac 1{i!}<\left.\frac {\part ^i\gamma (\lambda
{\hat{u}_N},\varepsilon  )}
{\part \varepsilon  ^i}\right|_{\varepsilon  =0},\eta _{N-i}>\, 
d\lambda\nonumber
\\&=&\frac 1 {N!}\left.\frac {\part ^N}{\part \varepsilon  ^N}
\right|_{\varepsilon  =0}
\int _0^1<\gamma (\lambda {\hat{u}_N},\varepsilon  ), {\hat{u}_N} >\,
d\lambda =\frac 1 {N!}\left.\frac
 {\part ^N \tilde{H}({\hat{u}_N},\varepsilon )}
{\part \varepsilon  ^N}\right|_{\varepsilon  =0}.
\nonumber 
\end{eqnarray}
The cotangent vector field $\hat {\gamma }_N$ is indeed a gradient field,
because we can show that 
\begin{equation}
\hat {\gamma }_N({\hat{\eta}_N} )=\frac {\delta (\textrm{per}_N\tilde{H})
}{\delta {\hat{\eta}_N} } ({\hat{\eta}_N} ).
\label{pvector}\end{equation}
According to Definition \ref{def:gradientfield} and using 
Lemma \ref{lemma:gateaxderivativeofptf},
for any $S_i\in T(M(\eta _i))$ we can compute that
\begin{eqnarray} &&<\frac {\delta }{\delta \eta   _i }\,\Bigl( \frac 1 {N!}
\left.\frac {\part ^N \tilde{H}({\hat{u}_N},\varepsilon  )}
{\part \varepsilon  ^N}\right|_{\varepsilon  =0}
\Bigr), S_i(\eta_i) >=\Bigl( \frac 1 {N!}
\left.\frac {\part ^N \tilde{H}({\hat{u}_N} ,\varepsilon )}
{\part \varepsilon  ^N}\right|_{\varepsilon  =0}
\Bigr)'(\eta _i)[S_i(\eta_i) ]\nonumber\\&=&
\frac 1 {N!}
\left.\frac {\part ^N }{\part \varepsilon  ^N}\right|_{\varepsilon  =0}
\tilde{H}'({\hat{u}_N},\varepsilon  )
[\varepsilon  ^iS_i(\eta_i) ]=\frac 1 {(N-i)!}
\left.\frac {\part ^{N-i} }{\part \varepsilon  ^{N-i}}
\right|_{\varepsilon  =0}\tilde{H}'({\hat{u}_N} ,\varepsilon )[S_i(\eta_i) ]
\nonumber\\
&=&\frac 1 {(N-i)!}\left.\frac {\part ^{N-i} }
{\part \varepsilon  ^{N-i}}\right|_{\varepsilon  =0}<\frac 
{\delta \tilde{H}}
{\delta {\hat{u}_N}}({\hat{u}_N},\varepsilon ), S_i(\eta_i) >=
\frac 1 {(N-i)!}\left.\frac {\part ^{N-i} }
{\part \varepsilon  ^{N-i}}\right|_{\varepsilon  =0}
<\gamma ({\hat{u}_N},\varepsilon  ), S_i(\eta_i) >
\nonumber\\
&=&<\frac 1 {(N-i)!}
\left.\frac {\part ^{N-i}\gamma ({\hat{u}_N},\varepsilon  ) }
{\part \varepsilon  ^{N-i}}\right|_{\varepsilon  =0}, 
S_i(\eta_i) >,\ 0\le i\le N.
\nonumber 
\end{eqnarray}
This equality implies that (\ref{pvector}) holds. It follows that 
the perturbation system (\ref{newhamiltonianstru}) is a Hamiltonian system.

Let us now turn to 
the property (\ref{Poissonpro}) for the Poisson bracket. 
Set $\gamma  _1 (\varepsilon)
 =\frac {\delta \tilde{H}_{1}}{\delta u}(\varepsilon),\gamma _2 (\varepsilon)
=\frac {\delta \tilde{H}_{2}}{\delta u}(\varepsilon)
\in T^*(M)$. In virtue of (\ref{pvector}), we can make the computation
\begin{eqnarray}&& \{ \textrm{per}_N\tilde{H}_{1},\textrm{per}_N
\tilde{H}_{2}\}
_{\hat{J}_{N}}(\hat{\eta}_{N})=<\frac {\delta (\textrm{per}_N
\tilde{H}_{1}) }{\delta \hat{\eta}_{N}}(\hat{\eta}_{N}),
\hat{J}_{N}(\hat{\eta}_{N})\frac {\delta (\textrm{per}_N
\tilde{H}_{2})}{\delta \hat{\eta}_{N}}(\hat{\eta}_{N})>
\nonumber\\&=&
\sum_{i=0}^N<\frac 1 {(N-i)!}\left.
\frac {\part ^{N-i}\gamma _1  (\hat {u}_N,\varepsilon )}{\part \varepsilon  ^
{N-i}}\right|_{\varepsilon  =0},\sum_{j=N-i}^N\frac 1 {(i+j-N)!}\times
\nonumber\\&&\qquad\quad
 \left.
\frac {\part ^{i+j-N}J(\hat {u}_N,\varepsilon)}{\part \varepsilon  ^
{i+j-N}}\right|_{\varepsilon  =0}\frac 1 {(N-j)!}
\left.\frac {\part ^{N-j}\gamma _2  (\hat {u}_N,\varepsilon )}
{\part \varepsilon  ^{N-j}}\right|_{\varepsilon  =0}>
\nonumber\\&=&
\sum_{i=0}^N<\frac 1 {(N-i)!}\left.
\frac {\part ^{N-i}\gamma _1  (\hat {u}_N,\varepsilon )}{\part \varepsilon  ^
{N-i}}\right|_{\varepsilon  =0}, \frac 1 {i!}\left.
\frac {\part ^{i}(J (\hat {u}_N,\varepsilon )
\gamma _2  (\hat {u}_N,\varepsilon ))}
{\part \varepsilon  ^{i}}\right|_{\varepsilon  =0}>
\nonumber\\&=&
\frac 1{N!}\left.\frac 
{\part ^N}{\part \varepsilon  ^N}\right|_{\varepsilon  =0}
<\gamma _1  (\hat {u}_N,\varepsilon ),J(\hat {u}_N,\varepsilon )
\gamma _2  (\hat {u}_N,\varepsilon )>=
(\textrm{per}_N \{\tilde{H}_1,\tilde{H}_2\}_J)(\hat{\eta}_{N}). \nonumber
\end{eqnarray}
This shows that the property (\ref{Poissonpro}) holds for the Poisson bracket.

Further, noting the structure of the perturbation Hamiltonian operators,
a multi-Hamiltonian formulation may readily be established for
the perturbation system. Therefore the proof is completed.
$\vrule width 1mm height 3mm depth 0mm$

\label{theory}

We should realize that two formulas (\ref{constantsformular}) and
(\ref{pcvf}) provide the explicit structures for 
the perturbation Hamiltonian functionals 
and the perturbation cotangent vector fields.
The whole theory above can be applied to all soliton hierarchies
and thus various interesting perturbation systems including
higher dimensional integrable couplings may be presented.
In the next section, we will however be only concerned with 
an application of the theory to the KdV soliton hierarchy.    

\section{Application to the KdV hierarchy}
\setcounter{equation}{0}

Let us consider the case of the KdV hierarchy
\begin{equation} u_t=K_n=K_n(u)=
(\Phi (u))^n u_x,\  \Phi=\Phi (u)=\part _x^2 + 2u_x\part _x^{-1}+4u,
\ n\ge 0. \label{KdVhierarchy}\end{equation}
Except the first linear equation $u_t=u_x$, each equation 
$u_t=K(u)$ $(n\ge 1)$ can be written as the following bi-Hamiltonian equation
\cite{Magri-JMP1978}
\begin{equation} u_t=K_n=J \frac {\delta {\tilde H}_{n} }{\delta u}=
M\frac {\delta {\tilde H}_{n-1} }{\delta u}. \end{equation} 
The corresponding Hamiltonian pair and Hamiltonian functionals read as
\begin{eqnarray}  && J=\part _x, \ M=M(u)=\part _x^3+2(\part _xu+u\part _x),
\label{HamiltonianpairforKdV}
\\ && {\tilde H}_n=\int H_n\,dx ,\
H_n= H_n(u)=\int_0^1uf_n(\lambda u)\, d\lambda,\ f_n=\Psi ^nu, \ n\ge 0,
\end{eqnarray} 
where $\Psi=\Phi ^\dagger=\part _x^2 +4u-2\part _x^{-1}u_x$. 
Therefore each equation in the KdV hierarchy (\ref{KdVhierarchy})
has infinitely many commuting symmetries $\{K_m\}_{m=0}^\infty$
and conserved densities $\{H_m\}_{m=0}^\infty$.

The second equation in the hierarchy (\ref{KdVhierarchy}) gives the 
following KdV equation
\begin{equation}  u_t=u_{xxx}+6uu_x, \label{KdVeq} \end{equation}
which serves as a well-known model of soliton phenomena. Its many
remarkable properties were reviewed by Miura \cite{Miura-SIAMR1976}.
In our discussion, we are concerned only with bi-Hamiltonian formulations
and consequent symmetries and conserved densities.
The bi-Hamiltonian formulation of the KdV equation (\ref{KdVeq}) 
can be written down  
\begin{equation}u_t=J\frac {\delta \tilde { H}_1}{\delta u}=M \frac 
{\delta \tilde {H}_0}{\delta u}
\label{biHamiltonianformulationofKdV}\end{equation} 
with
two Hamiltonian functionals
\begin{equation}
\tilde { H}_0=\int H_0\,dx =\int \frac 12 u^2\,dx,\ \tilde { H}_1=
\int H_1\,dx= \int (\frac 12 uu_{xx}+u^3)\, dx.
\end{equation}
It has also an isospectral zero curvature representation 
$U_t-V_x+[U,V]=0$ with
\begin{equation} U=\left[\ba {cc} 0& -u-\lambda 
\vspace{2mm} \\ 1&0 \ea \right],\
V=\left[\ba {cc} u_x&-u_{xx}-2u^2+2\lambda u+4\lambda ^2 \vspace{2mm} \\
2u-4\lambda & -u_x\ea \right],\label{zcrepofKdV}
\end{equation} 
where $\lambda $ is a spectral parameter (see \cite{AblowitzS-book1981}
for more information).  
These two properties will be used to construct bi-Hamiltonian formulations 
and zero curvature representations for the related perturbation systems. 

In order to apply the general idea of constructing integrable couplings
to the KdV equations, let us start from the following perturbed equation
\begin{equation}u_t=K^{\rm{per}}(u)=\sum_{i=0}^\infty \alpha _i
\varepsilon ^i S_i(u), \label{gKdVpie}\end{equation} 
where $\alpha _i $ are arbitrary constants and
the $S_i$ are taken from zero function and $K_n,\,n\ge 0$,
so that the series (\ref{gKdVpie}) terminates.
To obtain integrable couplings of the $n$-th order KdV equation $u_t=K_n$,
we need to fix $S_0=K_n$.
Various integrable couplings can be generated by making the perturbation
defined by (\ref{multiplescaleperturbationseries}). In what follows, 
we would only like to present some illustrative examples.

\subsection{Standard perturbation systems}

First of all, let us choose the $n$-th order KdV equation itself as
an initial equation:
\[ u_t=K^{\rm{per}}(u)=K_n(u) \]
for each $n\ge 1$.
In this case, the single scale perturbation $\hat{u}_N=\sum_{i=0}^N
\varepsilon ^i\eta _i(x,t)$ leads to a type of integrable couplings:
\begin{equation}\hat {\eta }_{Nt}=\hat {K}_{nN}(\hat {\eta }_N),\ N\ge 0,
\label{perturbationsystemofnthKdV}\end{equation} 
which are called the standard perturbation systems of $u_t=K_n$
and have been discussed in \cite{TamizhmaniL-JPA1983,MaF-PLA1996}.
These systems have the following bi-Hamiltonian formulations \cite{MaF-PLA1996}
\begin{equation}\hat {\eta }_{Nt}= \hat {K}_{nN}(\hat {\eta }_{N})=
\hat {\Phi }_{N}^n\hat {\eta }_{Nx}=
\hat {J }_{N}\frac {\delta ( \textrm{per}_N \tilde {{H}}
_{n})}{\delta \hat {\eta }_N}=
\hat {M }_{N}\frac {\delta (\textrm{per}_N \tilde {H}_{n-1})}
{\delta \hat {\eta }_N} ,\end{equation}
where the Hamiltonian functionals $\textrm{per}_N\tilde{H }_{n}$,
the hereditary recursion operator $\hat {\Phi }_{N}$ and
the Hamiltonian pair $\{\hat {J }_{N},\hat {M }_{N}\}$ are given by
\begin{eqnarray}  &&   \textrm{per}_N \tilde{H}_{n} = \frac 1{N!}\frac
{\part ^N \tilde {H}_n (\hat {u}_N)}{\part \varepsilon ^N },
\label{hat{H}{nN}} \\ &&
\hat {\Phi }_{N} = \left [
\begin{array} {cccc}  \Phi_0(\eta _0) & & & 0\vspace{2mm}\\
\Phi_1(\eta _1)&\Phi _0(\eta _0) & & \vspace{2mm}\\
\vdots & \ddots & & \vspace{2mm}\\
\Phi _N(\eta _N)& \cdots &\Phi _1(\eta _1)&\Phi _0(\eta _0)
\end{array} \right ],
 \\
&&  \hat {J }_{N}=
\left [ \begin{array} {cccc} 0& & & \part _x
\vspace{2mm}\\  & &\part _x & \vspace{2mm}\\
& \begin{turn} {45}\vdots \end{turn} & & \vspace{2mm}\\
\part _x & & & 0
\end{array} \right ],\ 
\hat {M }_{N} = \left [\begin{array} {cccc} 0& & & M_0(\eta _0)
\vspace{2mm}\\ &  &M_0(\eta _0)& M_1(\eta _1)\vspace{2mm}\\
& \begin{turn} {45}\vdots \end{turn}&  \begin{turn} {45}\vdots \end{turn}
& \vdots \vspace{2mm}\\
M_0(\eta _0) &M_1(\eta _1) & \cdots &M_N(\eta _N) 
\end{array} \right ], \qquad \ \, 
\end{eqnarray} 
with
\begin{equation}M_i=M_i(\eta _i)=\delta_{i0}\part _x^3
+2(\part _x\eta _i+\eta _i\part _x),
\ \Phi _i=
\Phi_i(\eta _i)=\delta_{i0}\part _x^2+2(\part _x\eta _i\part _x^{-1}
+\eta _i) ,\ 0\le i\le N . \label{Mi(etai)andPhii(etai)}\end{equation} 
Moreover they have infinitely many commuting symmetries 
$\{\hat {K}_{mN}\}_{m=0}^\infty$ and 
conserved densities $\{\hat {H}_{mN}\}_{m=0}^\infty$.

We list the first two standard perturbation systems of 
the KdV equation (\ref{KdVeq}):
\begin{eqnarray}  &&\left \{\begin{array} {l} 
\eta _{0t}=\eta _{0xxx}+6\eta _{0}\eta _{0x}, \vspace{2mm}\\
\eta _{1t}=\eta _{1xxx}+6(\eta _{0}\eta _{1})_x;\end{array} \right.
\label{firstorderpsofKdV}\\ && 
\left \{\begin{array} {l} 
\eta _{0t}=\eta _{0xxx}+6\eta _{0}\eta _{0x}, \vspace{2mm}\\
\eta _{1t}=\eta _{1xxx}+6(\eta _{0}\eta _{1})_x,\vspace{2mm} \\
\eta _{2t}=\eta _{2xxx}+6\eta _{1}\eta _{1x}+6(\eta _{0}\eta _{2})_x.
\end{array} \right. \label{secondorderpsofKdV}
\end{eqnarray} 
The first-order perturbation system (\ref{firstorderpsofKdV}) has 
the following bi-Hamiltonian formulation 
\begin{equation}
\hat{\eta}_{1t} = \left[  \begin{array} {cc} 0&\part _x \vspace{2mm}\\ 
\part _x&0 \end{array} \right]
\frac {\delta (\textrm{per}_1 \tilde {H}_{1})} {\delta \hat {\eta} _1}=
\left[  \begin{array} {cc} 0 &\part _x^3 +2\eta _{0x}+4\eta _0\part _x
 \vspace{2mm}\\ \part _x^3 +2\eta _{0x}+4\eta _0\part _x
& 2\eta _{1x}+4\eta _1\part _x \end{array} \right]
\frac {\delta  (\textrm{per}_1\tilde{H}_{0})}{\delta \hat {\eta} _1}
\end{equation}
with $\hat {\eta} _1= (\eta _0,\eta _1)^T$
and the Hamiltonian functionals
\begin{eqnarray}  &&
\textrm{per}_1\tilde{H}_{0}=\int \hat {H}_{01}\,dx,\ 
\hat {H}_{01}=\eta _0\eta _1,
\nonumber \\ && \textrm{per}_1\tilde{H}_{1}
=\int \hat {H}_{11}\,dx,\
\hat {H}_{11}=
\frac 12 (\eta _0 \eta _{1xx}+\eta _{0xx}\eta _1)+3\eta _0^2\eta _1.
\nonumber \end{eqnarray}  
The second-order perturbation system (\ref{secondorderpsofKdV}) has 
the following bi-Hamiltonian formulation
\begin{equation}
 \hat {\eta }_{2t}=\hat {J}_2
\frac {\delta (\textrm{per}_2 \tilde{H}_{1})}{\delta \hat {\eta  }_2}=
\hat{M}_2
\frac {\delta (\textrm{per}_2\tilde{H}_{0})}{\delta \hat {\eta  }_2},\ 
\hat{\eta} _2 =(\eta _0,\eta _1,\eta _2)^T
\end{equation} 
with the Hamiltonian functionals
\begin{eqnarray}   &&\textrm{per}_2\tilde{H}_{0}=\int \hat {H}_{02}\,dx,\ 
\hat H_{02}= \eta  _0\eta  _2+\frac 12 \eta  _1^2 ,
\nonumber \\&& \textrm{per}_2\tilde{H}_{1}=\int \hat {H}_{12}\,dx,\ 
 \hat H_{12}=
 \frac 12 (\eta  _0\eta  _{2xx}+\eta  _1\eta  _{1xx}+\eta  _{0xx}\eta  _2)+3
\eta  _0\eta  _1^2+3\eta  _0^2\eta  _2.\nonumber
\end{eqnarray}  

Another example that we want to show is the first-order standard 
perturbation system of the fifth order KdV equation $u_t=K_2(u)$:
\begin{equation}  
\left\{ \begin{array} {l} 
\eta _{0t}=\eta _{0,5x}+10\eta _0\eta _{0xxx}+20\eta _{0x}\eta _{0xx}
+30\eta _0^2\eta _{0x},
\vspace{2mm}\\
\eta _{1t}=\eta _{1,5x}+10 \eta _{0xxx}\eta _1+10 \eta _0\eta _{1xxx}+
20\eta _{0xx}\eta _{1x}+20\eta _{0x}\eta _{1xx}+60\eta _{0}\eta _{0x}\eta _1
+30 \eta _0^2\eta _{1x},
\end{array} \right.
\end{equation}
where $\eta _{0,5x}$ and $\eta _{1,5x}$, as usual, stand for 
the fifth order derivatives of $\eta _0$ and $\eta _1$ with respect to $x$.
It has the following bi-Hamiltonian formulation 
\begin{equation}  \hat{\eta }_{2t}=\hat {J}_1\frac {\delta 
(\textrm{per}_1\tilde{H}_2)}{\delta \hat{\eta }_1}=
\hat {M}_1\frac {\delta 
(\textrm{per}_1\tilde{H}_1)}{\delta \hat{\eta }_1},
 \end{equation}
where the Hamiltonian function $\textrm{per}_1\tilde{H}_1$
is given as before and the Hamiltonian function 
 $\textrm{per}_1\tilde{H}_2$, 
given as follows 
\begin{equation} 
\textrm{per}_1\tilde{H}_2   = \frac12 \eta _{0xxxx}\eta _1+\frac1 2\eta _0
\eta _{1xxxx}+\frac {20}3 \eta _0\eta _{0xx}\eta _1 +\frac {10}3 
\eta _0^2 \eta _{1xx}+\frac 53 \eta _{0x}^2 \eta _1+
\frac {10}3 \eta _0\eta _{0x}\eta _{1x} +\frac {40}3 \eta _{0}^3 \eta _1.
\nonumber 
 \end{equation}

\subsection{Nonstandard perturbation systems}

Secondly, let us choose
a perturbed equation 
\[ u_t=K^{\rm{per}}(u,\varepsilon)=K_n+\alpha \varepsilon K_n, \ 
\alpha =\textrm{const.},\  \alpha \ne 0,\]
as an initial equation for each $n\ge 1$.
This equation can be viewed as
\begin{eqnarray}   u_t=K^{\rm{per}}(u,\varepsilon)& =&
J \frac {\delta ({\tilde {H}}_n
+ \alpha \varepsilon  {\tilde {H}}_{n}) }{\delta u}   =
M \frac {\delta ({\tilde {H}}_{n-1}
+ \alpha \varepsilon  \tilde {H}_{n-1}) }{\delta u}  \nonumber \\ &=&
(J+ \alpha \varepsilon J) \frac { \delta {\tilde {H}} _{n} }{\delta u}
=  (M+ \alpha \varepsilon M) \frac {\delta \tilde  {H}_{n-1} }{\delta u}
.\end{eqnarray} 
Therefore the corresponding perturbation systems also
have quadruple Hamiltonian formulations. We focus on the 
first-order perturbation system under the single scale perturbation.
It has the quadruple Hamiltonian formulation
\begin{equation}\hat {\eta }_{1t} =
\hat {J}_{1}^{(1)}\frac {\delta (\textrm{per}_1\tilde {H}_{n}^{(1)})}
{\delta \hat{\eta }_{1}}
=\hat {M}_{1}^{(1)}\frac {\delta (\textrm{per}_1\tilde{H}_{n-1}^{(1)})}
{\delta \hat{\eta }_{1}}
=\hat {J}_{1}^{(2)}\frac {\delta (\textrm{per}_1\tilde{H}_{n}^{(2)})}
{\delta \hat{\eta }_{1}}
=\hat {M}_{1}^{(2)}\frac {\delta (\textrm{per}_1\tilde {H}_{n-1}^{(2)})}
{\delta \hat{\eta }_{1}},   \label{nonstandardperturbationsystemofnthKdV}
\end{equation} 
namely,
\begin{eqnarray}  
\hat {\eta }_{1t} & = &
\left[ \begin{array} {cc } 0& \part \vspace{2mm}\\ \part & 0 \end{array} 
\right ]
\frac{\delta (\textrm{per}_1\tilde{ {H}}_{n}+ \alpha 
\tilde{ {H}}_{n}(\eta _0))}
{\delta \hat {\eta }_1  }
=\left[ \begin{array} {cc } 0& M_0 \vspace{2mm}\\ M_0 & M_1 \end{array} 
\right ]
\frac{\delta (\textrm{per}_1\tilde{{H}}_{n-1}+ \alpha \tilde{{H}}_{n-1}
(\eta _0))} {\delta \hat {\eta }_1 }
\nonumber \\ &= &
\left[ \begin{array} {cc } 0& \part \vspace{2mm}\\ \part & \alpha \part 
\end{array} \right ]
\frac{\delta  (\textrm{per}_1\tilde{{H}}_{n})}{\delta \hat {\eta }_1} 
=\left[ \begin{array} {cc } 0& M_0 \vspace{2mm}\\ M_0 & M_1+\alpha M_0 
\end{array} \right ]
\frac{\delta (\textrm{per}_1\tilde{{H}}_{n-1})}{\delta  \hat {\eta }_1 },
\label{11perturbationsystem}\end{eqnarray} 
where the functionals $ \textrm{per}_1\tilde{{H}}_{n},\,
\textrm{per}_1\tilde{{H}}_{n-1}$ and the operators
$M_i$ are defined by (\ref{hat{H}{nN}}) and
(\ref{Mi(etai)andPhii(etai)}), respectively.
Since two Hamiltonian operators $\hat {J}_{1}^{(1)}$
and $\hat {J}_{1}^{(2)}$ are invertible, we can obtain five hereditary
recursion operators for the equation
(\ref{nonstandardperturbationsystemofnthKdV}):
\begin{eqnarray} 
&& \hat {J}_{1}^{(2)} (\hat {J}_{1}^{(1)} )^{-1}=\left[\begin{array}
 {cc}
1&0 \vspace{2mm}\\ \alpha & 1 \end{array} \right],\
\hat {J}_{1}^{(1)} (\hat {J}_{1}^{(2)} )^{-1}=\left[\begin{array} {cc}
1&0 \vspace{2mm}\\ -\alpha & 1 \end{array} \right],  \nonumber \\ &&
\hat {M}_{1}^{(1)} (\hat {J}_{1}^{(2)} )^{-1}=\left[\begin{array} {cc}
\Phi _0  &0 \vspace{2mm}\\ \Phi _1-\alpha \Phi _0& \Phi _0 \end{array} 
\right],\
\hat {M}_{1}^{(2)} (\hat {J}_{1}^{(1)} )^{-1}=\left[\begin{array} {cc}
\Phi _0  &0 \vspace{2mm}\\ \Phi _1+\alpha \Phi _0& \Phi _0 \end{array} 
\right],
\nonumber \\ && 
\hat {M}_{1}^{(1)} (\hat {J}_{1}^{(1)} )^{-1}
=\hat {M}_{1}^{(2)} (\hat {J}_{1}^{(2)} )^{-1}
=\left[\begin{array} {cc}
\Phi _0  &0 \vspace{2mm}\\ \Phi _1 & \Phi _0 \end{array} \right],
\nonumber  \end{eqnarray}
where the operators $\Phi_i$ are defined by (\ref{Mi(etai)andPhii(etai)}).
These operator structures suggest two classes of hereditary
recursion operators for the equation
(\ref{nonstandardperturbationsystemofnthKdV})
\begin{equation}
\hat {\Phi }_{1}^{(1)}(\beta )=\left[\begin{array} {cc}
\beta _0 &0 \vspace{2mm}\\ \beta _1& \beta _0 \end{array} \right],\
\hat {\Phi }_{1}^{(2)}(\beta )=\left[\begin{array} {cc}
\beta _0\Phi _0&0 \vspace{2mm}\\ \beta _0\Phi _1 +\beta _1\Phi _0& \beta _0
\Phi _0 \end{array} \right],
\label{tworos} \end{equation} 
where $\beta =(\beta _0,\beta _1)^T$ with the $\beta _i$ being
arbitrary constants.
They are really hereditary operators and recursion operators
for the equation (\ref{nonstandardperturbationsystemofnthKdV}),
which can be verified by direct computation or by viewing them as
the first-order perturbation operators of the initial operators $\beta _0
+\beta _1\varepsilon$ and $\beta _0\Phi +\beta _1 \varepsilon \Phi$.  
Therefore the integrable coupling
(\ref{nonstandardperturbationsystemofnthKdV})
of the $n$-th order KdV equation $u_t=K_n(u)$ possesses 
two classes of hereditary recursion operators defined by (\ref{tworos}). 
These two classes of operators have the property 
\begin{equation}{\Phi }_{1}^{(1)}(\beta ){\Phi }_{1}^{(2)}(\gamma ) =
{\Phi }_{1}^{(2)}(\beta ){\Phi }_{1}^{(1)}(\gamma )=
\left[\begin{array} {cc}
\beta _0\gamma _0 \Phi _0&0 \vspace{2mm}\\ \beta _0\gamma _0\Phi _1 
+(\beta _0\gamma _1+\beta _1\gamma _0)\Phi _0&
\beta _0\gamma _0\Phi _0 \end{array} \right],
\end{equation} 
for any two constant vectors $\beta =(\beta _0,\beta _1)^T$ and $\gamma =
(\gamma _0,\gamma _1)^T$, which also shows that their product 
can not constitute completely new recursion operators.

We can also start from the perturbed KdV type equation
\begin{equation}u_t=K^{\textrm{per}}
(u,\varepsilon)= K_n+\al \varepsilon ^jK_{i_j},\ 
\al =\textrm{const.}, \ \alpha \ne 0,
  \end{equation} 
where $i_j$ is a natural number.
Let us illustrate the idea of construction by the following specific example
\begin{equation}u_t=K^{\textrm{per}}(u,\varepsilon)
= K_n + \al \varepsilon ^2 K_{n+1}\ (n\ge 1),
 \label{K_n+alphavarepsilonK_{n+1}} 
\end{equation}
which can be viewed as a tri-Hamiltonian system:
\begin{equation}
 u_t=K^{\textrm{per}} =J\frac {\delta (\tilde{H}_n+\alpha \varepsilon ^2
\tilde {H}_{n+1}) }{\delta u}
= (J+\alpha \varepsilon ^2M)\frac {\delta \tilde{H}_n }{\delta u}
=M\frac {\delta (\tilde{H}_{n-1}+\alpha \varepsilon ^2
\tilde {H}_{n}) }{\delta u}.
\end{equation}
Therefore, according to Theorem 
\ref{thm:perturbationbi-Hamiltonianformulation}, 
the second-order perturbation system of the the perturbed system 
(\ref{K_n+alphavarepsilonK_{n+1}})
\begin{equation}\left \{  
\begin{array} {l} \eta _{0t } = {K}_{n}(\eta _0), \vspace{2mm}\\
\eta _{1t} =K_n'(\eta _0)[\eta _1], \vspace{2mm}\\
 \eta _{2t}= \frac 12\left.\frac
 {\part ^2 K_n(\hat{u}_2) }{\part \varepsilon ^2}
 \right|_{\varepsilon =0}+ \alpha K_{n+1}(\eta _0)\ea
\right. \label{biggerK_npe} \end{equation} 
possesses the following tri-Hamiltonian formulation
\begin{equation}
\hat {\eta }_{2t}=
\hat {J}_2^{(1)}\frac {\delta \tilde{{H}}_n^{(1)}}{\delta \hat {\eta  }_2}=
\hat{J}_2^{(2)}\frac {\delta \tilde{{H}}_n^{(2)}}{\delta \hat {\eta  }_2}=
\hat{J}_2^{(3)}
\frac {\delta \tilde{{H}}_n^{(3)}}{\delta \hat {\eta  }_2},\
\hat{\eta} _2=(\eta _0,\eta _1,\eta _2)^T
\end{equation} 
with a triple of Hamiltonian operators 
\begin{equation}
\hat {J}_2^{(1)}=
\left[\ba {ccc}0&0&\part _x \vspace{2mm}\\ 0&\part _x& 0\vspace{2mm}\\
\part _x& 0& 0\ea \right],\ 
\hat {J}_2^{(2)}=
\left[\ba {ccc}0&0&\part _x \vspace{2mm}\\ 0&\part _x& 0\vspace{2mm}\\
\part _x& 0& \alpha M_0\ea \right],\  
\hat {J}_2^{(3)}=
\left[\ba {ccc}0&0&M_0 \vspace{2mm}\\ 0&M_0& M_1\vspace{2mm}\\
M_0 & M_1& M_2\ea \right] 
\end{equation}
and the corresponding three Hamiltonian functionals
\begin{equation} \left\{ \begin{array}{l}
\tilde{H}_n^{(1)} (\hat{\eta } _2)= (\textrm{per}_2\tilde{{H}}_{n})
(\hat {\eta }_2)+ \alpha \tilde {H}_{n+1}(\eta _0),\vspace{2mm} \\
 \tilde{H}_n^{(2)} (\hat{\eta } _2)= (\textrm{per}_2\tilde{{H}}_{n})
(\hat {\eta }_2),\vspace{2mm}\\ 
 \tilde{H}_n^{(3)} (\hat{\eta } _2)= (\textrm{per}_2\tilde{{H}}_{n-1})
(\hat {\eta }_2)+ \alpha \tilde {H}_{n}(\eta _0).
 \end{array}\right. \end{equation}
Similarly, the perturbation system (\ref{biggerK_npe}) 
has also two classes of hereditary recursion operators: 
\begin{equation}
\hat {\Phi }_2^{(1)}(\beta )= \left[ \begin{array}{ccc} \beta _0& 0& 0 \vspace{2mm}\\ 
\beta _1& \beta_0&  \vspace{2mm}\\ 
\beta _2& \beta _1& \beta_0 \end{array} \right],\ 
\hat {\Phi }_2^{(2)}(\beta )
= \left[ \begin{array}{ccc} \beta _0\Phi _0& 0& 0 \vspace{2mm}\\ 
\beta _0\Phi _1+\beta _1\Phi _0 & \beta_0\Phi _0&  \vspace{2mm}\\ 
\beta _0\Phi _2+\beta _1\Phi _1+\beta _2\Phi _0
& \beta _0\Phi _1+\beta _1\Phi _0 & \beta_0\Phi _0 \end{array} \right],
\end{equation}
where the operators $\Phi_i$ are defined by (\ref{Mi(etai)andPhii(etai)})
and $\beta =(\beta _0,\beta _1,\beta _2)^T$ is a constant vector.

Let us fix $n=1$ and then the system (\ref{biggerK_npe}) gives 
an integrable coupling of the KdV equation (\ref{KdVeq}),
which possesses the following tri-Hamiltonian formulation
\begin{equation}
\hat {\eta }_{2t}=
\hat {J}_2^{(1)}\frac {\delta \tilde{{H}}_1^{(1)}}{\delta \hat {\eta  }_2}=
\hat{J}_2^{(2)}\frac {\delta \tilde{{H}}_1^{(2)}}{\delta \hat {\eta  }_2}=
\hat{J}_2^{(3)}
\frac {\delta \tilde{{H}}_1^{(3)}}{\delta \hat {\eta  }_2}
\end{equation}
with three Hamiltonian functionals
\begin{equation}  \left\{ \begin{array} {l}
\tilde{H}_1^{(1)} (\hat{\eta } _2)=
 \frac 12 (\eta  _0\eta  _{2xx}+\eta  _1\eta  _{1xx}+\eta  _{0xx}\eta  _2)+3
\eta  _0\eta  _1^2+3\eta  _0^2\eta  _2
 \vspace{2mm} \\ \qquad \qquad\quad 
+\alpha (\frac12 \eta _0\eta _{0xxxx}+\frac {10}3\eta _0^2\eta _{0xx}
+\frac 53 \eta _0\eta _{0x}^2+\frac {10}3 \eta _0^4),
 \vspace{2mm} \\
 \tilde{H}_1^{(2)} (\hat{\eta } _2)=
 \eta _0\eta _2+\frac12 \eta _1^2
 +\alpha (\frac12 \eta _0\eta _{0xx}+\eta _0^3),
 \vspace{2mm} \\
 \tilde{H}_1^{(3)} (\hat{\eta } _2)=
 \frac 12 (\eta  _0\eta  _{2xx}+\eta  _1\eta  _{1xx}+\eta  _{0xx}\eta  _2)+3
\eta  _0\eta  _1^2+3\eta  _0^2\eta  _2.
\end{array}  \right.  \end{equation}

In order to distinguish the standard perturbation systems
defined by (\ref{perturbationsystemofnthKdV}),
the integrable couplings of the $n$-th order KdV equation $u_t=K_n$, defined by
(\ref{nonstandardperturbationsystemofnthKdV}) and (\ref{biggerK_npe}),
are called the non-standard perturbation systems.
Interestingly, each of these systems has both a local multi-Hamiltonian 
formulation and two classes of hereditary recursion operators.

\subsection{2+1 dimensional integrable couplings}

Thirdly, let us consider a case of bi-scale perturbations
(\ref{specificcaseofperturbation}), i.e., 
\begin{equation}\hat {u}_N
=\sum_{i=0}^N\varepsilon ^i \eta _{i}, \ \eta _i=\eta _i(x,y,t), \ 
y=\varepsilon x. \nonumber
\end{equation} 
In order to present explicit results for integrable couplings, 
we take the KdV equation
(\ref{KdVeq}) as an illustrative example, due to its simplicity.
We recall that the KdV equation (\ref{KdVeq})
has the bi-Hamiltonian formulation (\ref{biHamiltonianformulationofKdV})
and the Lax pair (\ref{zcrepofKdV}). 

Let us introduce the bi-scale perturbation series above 
into the KdV equation (\ref{KdVeq}) and equate powers of $\varepsilon $. 
As a $N$-th order approximation, 
we obtain a $2+1$ dimensional perturbation systems of evolution equations
\begin{equation} 
\left \{
\begin{array}{l}
\eta  _{0t_1}=\eta  _{0xxx}+6\eta  _0\eta  _{0x},\\
\eta  _{1t_1}=\eta  _{1xxx}+3\eta  _{0xxy}+6(\eta  _0\eta  _{1})_x
+6\eta  _0\eta  _{0y},\\
\eta  _{2t_1}=\eta  _{2xxx}+3\eta  _{1xxy}+3\eta  _{0xyy}+6(\eta  _0
\eta  _{2})_x
+6\eta  _1\eta  _{1x}+6(\eta  _0\eta  _{1})_y,\\
\eta  _{jt_1}=\eta  _{jxxx}+3\eta  _{j-1,xxy}+3\eta  _{j-2,xyy}
+\eta  _{j-3,yyy}
\\ \ \qquad +6\Bigl(\sum_{i=0}^j\eta  _i\eta  _{j-i,x}+\sum_{i=0}^{j-1}
\eta  _i\eta  _{j-i-1,y}
\Bigr),\ 3\le j\le N.\end{array}\right.
\label{2+1KdVperturbationsystem}\end{equation} 
This system has been already presented in \cite{MaF-PLA1996}.
It follows from our general theory that it gives an integrable coupling
of the KdV equation (\ref{KdVeq}).

In what follows,
we would like to propose a bi-Hamiltonian formulation and the consequent
hereditary recursion operator for the system
(\ref{2+1KdVperturbationsystem}). To the end,
we first need to compute a perturbation Hamiltonian pair by Theorem
\ref{thm:perturbationHamiltonianoperator}:
\begin{eqnarray}  
&& \hat{J}_N = \left [ \begin{array} {ccccc} 0 & & &  & \part _x \\
&  & & \part _x &\part _y \\
  & & \part _x &\part _y &0 \\
&  \begin{turn}{45}\vdots\end{turn} 
&  \begin{turn}{45}\vdots\end{turn} &
 \begin{turn}{45}\vdots\end{turn}& \vdots \\
\part _x & \part _y  & 0 &  \cdots & 0
 \end{array} \right], \label{2+1firstHamiltonianoperator} \\
&&  \hat{M}_N = \left[ \begin{array} {ccccc}  
0 & & & & P(\varepsilon )|_{\varepsilon =0}\vspace{2mm} \\
 & & &  P(\varepsilon )|_{\varepsilon =0}&\frac {1}{1!}\left.
\frac {\part P(\varepsilon )}{\part \varepsilon }\right.\Bigl.\Bigr
|_{\varepsilon =0}\vspace{2mm} \\
& &  P(\varepsilon )|_{\varepsilon =0}&\frac {1}{1!}\left.
\frac {\part P(\varepsilon )}{\part \varepsilon }\right.\Bigl.\Bigr
|_{\varepsilon =0}&
\frac {1}{2!}\left.
\frac {\part ^2P(\varepsilon )}{\part \varepsilon ^2}
\right.\Bigl.\Bigr |_{\varepsilon =0}\vspace{2mm} \\
&\begin{turn}{45}\vdots\end{turn}
& \begin{turn}{45}\vdots\end{turn}  &\begin{turn}{45}\vdots\end{turn}
 &\vdots \vspace{2mm}  \\
 P(\varepsilon )|_{\varepsilon =0}&
 \frac {1}{1!}\left.
\frac {\part P(\varepsilon )}{\part \varepsilon }\right.\Bigl.\Bigr
|_{\varepsilon =0}
& \frac {1}{2!}\left.
\frac {\part ^2P(\varepsilon )}{\part \varepsilon ^2}\right.\Bigl.\Bigr
|_{\varepsilon =0}
 & \cdots & \frac {1}{N!}\left.\frac {\part ^N P(\varepsilon )}
{\part \varepsilon ^N}\right.\Bigl.\Bigr |_{\varepsilon =0} 
 \end{array} \right) ,\qquad\quad
 \label{2+1secondHamiltonianoperator}
  \end{eqnarray} 
where the differential operator $P(\varepsilon )$ represents  
\[P(\varepsilon )=(\part _x+\varepsilon \part _y)^3 +2[(
\part _x+\varepsilon \part _y)\hat {u}_N+\hat {u}_N(\part _x
+\varepsilon \part _y)].\]
The explicit expressions for various derivatives
of $P(\varepsilon )$ with respect to $\varepsilon $ 
can be obtained as follows:
\begin {equation}\left \{\begin{array}{l}
P(\varepsilon )|_{\varepsilon =0} =\part _x ^3+2(\part _x\eta _0+\eta _0
\part _x), \vspace{2mm}\\ 
 \frac {1}{1!}\left. 
\frac {\part P(\varepsilon )} {\part \varepsilon }\right.\Bigl.\Bigr
|_{\varepsilon =0}
=3\part _x ^2\part _y +2(\part _x\eta _1+\eta _1\part _x)+
2(\part _y\eta _0+\eta _0\part _y), \vspace{2mm}\\ 
 \frac {1}{2!}\left.\frac {\part ^2P(\varepsilon )}
{\part \varepsilon ^2}\right.\Bigl.\Bigr |_{\varepsilon =0}
=3\part _x \part _y ^2+2(\part _x\eta _2+\eta _2
\part _x) +2(\part _y\eta _1+\eta _1\part _y), \vspace{2mm}\\ 
 \frac {1}{3!}\left . \frac {\part ^3 
P(\varepsilon )}{\part \varepsilon ^3}\right.\Bigl.\Bigr |_{\varepsilon =0}
=\part _y ^3 +2(\part _x\eta _3+\eta _3
\part _x) +2(\part _y\eta _2+\eta _2\part _y), \vspace{2mm}\\ 
 \frac {1}{i!}
\left. \frac {\part ^iP (\varepsilon )}
{\part \varepsilon ^i}\right.\Bigl.\Bigr |_{\varepsilon =0}
=2(\part _x\eta _i+\eta _i \part _x) +2(\part _y\eta _{i-1}
+\eta _{i-1}\part _y), \ 4\le i\le N,
\end{array}\right. \end{equation} 
which gives rise to an explicit expression for the Hamiltonian operator
$\hat{M}_N$. Secondly, we need to compute the Hamiltonian functionals for 
the system (\ref{2+1KdVperturbationsystem}). Note that
$\part _x \to \part _x+\varepsilon \part _y$, and thus,
under the perturbation (\ref{specificcaseofperturbation}), we have 
\[u_{xx}\to \sum_{i=0}^N \varepsilon ^i(\eta _{ixx}+2\varepsilon \eta _{ixy}
+\varepsilon ^2\eta _{iyy}).\]
Further, by Theorem \ref{thm:perturbationbi-Hamiltonianformulation},
we obtain two perturbation Hamiltonian functionals:
\begin{eqnarray}    \textrm{per}_N\tilde{H}_{0}& =&
\iint \frac {1}{N!}  \frac {\part ^N {H}_0(\hat {u}_N)}
{\part \varepsilon ^N}\Bigl.\Bigr |_{\varepsilon =0} \,dxdy
=\iint \frac12 \sum_{i=0}^N\eta _i \eta _{N-i}\, dxdy,
\label{2+1firstHamiltonianfunctional} \\ 
 \textrm{per}_N \tilde {{H}}_{1}&=&
 \iint\frac {1}{N!}\frac {\part ^N {H}_1(\hat {u}_N)}
 {\part \varepsilon ^N} \Bigl.\Bigr |_{\varepsilon =0} \,dxdy
 =\iint \bigl[\frac12 \sum_{i+j=N}\eta _i \eta _{jxx} \bigr .\nonumber \\
 && \bigl. +\sum_{i+j=N-1}\eta _i\eta _{jxy}+
\frac12 \sum_{i+j=N-2}\eta _i\eta _{jyy}
+\sum_{i+j+k=N}\eta _i\eta _j\eta _k \bigr ]\, dxdy.
\label{2+1secondHamiltonianfunctional}
\end{eqnarray} 
Now the following bi-Hamiltonian formulation for the system
(\ref{2+1KdVperturbationsystem}) becomes clear:
\begin{equation} \hat {\eta }_{Nt}
=\hat {J}_N \frac {\delta (\textrm{per}_N\tilde{H}_1)}{\delta \hat{\eta }_N}
=\hat {M}_N \frac {\delta (\textrm{per}_N\tilde{H}_0)}{\delta \hat{\eta }_N},
 \label{2+1bi-Hamiltoniansystem}\end{equation} 
where $\hat {J}_N,\hat {M}_N,\textrm{per}_N\tilde{H}_0$ and $
\textrm{per}_N\tilde{H}_1$ are defined by
(\ref{2+1firstHamiltonianoperator}), (\ref{2+1secondHamiltonianoperator}),
(\ref{2+1firstHamiltonianfunctional}) and
(\ref{2+1secondHamiltonianfunctional}), respectively.
It should be realized that the $2+1$ dimensional bi-Hamiltonian system
(\ref{2+1bi-Hamiltoniansystem}) is local, because the Hamiltonian 
pair $\{\hat{J}_N,\hat{M}_N\}$ involves 
only the differential operators $\part _x$ and $\part _y$.

Theorem \ref{thm:perturbationrecursionoperator} guarantees the existence
of a hereditary recursion operator for the system
(\ref{2+1KdVperturbationsystem}).
It is of interest to get its explicit expression.
Note that the first Hamiltonian operator $\hat {J}_N$ has an invertible
operator
\begin{equation}  (\hat {J}_N)^{-1}=\left[ \begin{array} {cccc} P_N &P_{N-1}& \cdots
& P_0 \vspace{2mm}\\
P_{N-1}& &\begin{turn}{45}\vdots\end{turn} & \vspace{2mm}\\
\vdots &\begin{turn}{45}\vdots\end{turn}  & & \vspace{2mm}\\
P_0& & &0 \end{array} \right], \end{equation} 
where the operators $P_i$ are defined by
\begin{equation}  
P_0=\part _x^{-1},\ P_{1}=-\part _x^{-2}\part _y,\ \cdots,\ P_i=
(-1)^{i}\part _x^{-i-1}\part _y^{i} ,\ \cdots, \
P_N=(-1)^N\part _x^{-N-1}\part _y ^N.
\end{equation}
Therefore, the corresponding hereditary recursion operator is determined 
by $\hat {\Phi }_N=\hat {M}_N\hat {J}_N^{-1}$, but it can also be computed 
directly by Theorem \ref{thm:perturbationhereditaryoperator}:
\begin{equation}
\hat{\Phi }_N = \left[ \begin{array} {cccc}  
\Phi(\hat{u}_N)|_{\varepsilon =0}
 & & &  0 \\
\frac 1 {1!}\frac {\part \Phi(\hat{u}_N)}{\part \varepsilon}
\Bigl.\Bigr|_{\varepsilon =0} &\Phi(\hat{u}_N)|_{\varepsilon =0} &  & \\
\vdots & \ddots & \ddots & \\
\frac 1 {N!}\frac {\part ^N\Phi(\hat{u}_N)}{\part \varepsilon ^N}
\Bigl.\Bigr|_{\varepsilon =0} 
 &\cdots &  \frac 1 {1!}\frac {\part \Phi(\hat{u}_N)}{\part \varepsilon}
\Bigl.\Bigr|_{\varepsilon =0} & \Phi(\hat{u}_N)|_{\varepsilon =0} \ea
\right ]. \label{2+1dimensionalrecursionoperator}
\end{equation}
Here the operator $\Phi(\hat{u}_N)$ is defined by
\[\Phi(\hat{u}_N)= (\part _x+\varepsilon \part _y)^2+2(\hat {u}_{Nx}+
\varepsilon \hat {u}_{Ny})(\part _x+\varepsilon \part _y)^{-1} +
4\hat {u}_N,\]
and thus its $N+1$ derivatives with respect to $\varepsilon $ are found to be 
\begin{equation}  \left\{ \begin{array} {l} 
\Phi (\hat {u}_N) \bigl.\bigr|_{\varepsilon =0}=
\part _x ^2+2\eta _{0x}\part _x^{-1}+4\eta _0,
\vspace{2mm}  \\   
\frac 1 {1!}\frac {\part  \Phi (\hat {u}_N) }{\part
\varepsilon } \Bigl.\Bigr|_{\varepsilon =0}=
2\part _x\part _y+2(\eta _{1x}+\eta _{0y})\part _x^{-1}
-2\eta _{0x}\part _x^{-2}\part _y+4\eta _1,
\vspace{2mm} \\ 
\frac 1 {2!}\frac {\part ^2 \Phi (\hat {u}_N) }{\part
\varepsilon ^2} \Bigl.\Bigr|_{\varepsilon =0}=
\part _y^2+2(\eta _{2x}+\eta _{1y})\part _x^{-1}
-2(\eta _{1x}+\eta _{0y})\part _x^{-2}\part _y
+2\eta _{0x}\part _x^{-3}\part _y^2+4\eta _2,
\vspace{2mm} \\ 
\frac 1 {k!}\frac {\part  ^k\Phi (\hat {u}_N) }{\part
\varepsilon ^k} \Bigl.\Bigr|_{\varepsilon =0}=
\sum_{i+j=k}(-1)^j(2\eta _{ix}+\eta _{i-1,y})
\part _x^{-j-1}\part _y^j+4\eta_k,
\ 3\le k\le N,\end{array}\right. 
\label{concreteexpressionforderivativeofhatPhiN}
\end{equation} 
where we accept $\eta _{-1}=0$.

Let us now show the corresponding zero curvature representation
for the $2+1$ dimensional perturbation 
system (\ref{2+1KdVperturbationsystem}). By Theorem \ref{thm:zcrepofpe}
or (\ref{specificperturbationzcrep}),
the zero curvature representation for the system 
(\ref{2+1KdVperturbationsystem}) can be given by 
\be
\hat {U}_{Nt}-\hat {V}_{Nx}-\Pi \hat{V}_{Ny}+[\hat {U}_N,\hat {V}_N]=0,
\label{zcrepof2+1perturbationsystem} \end{equation} 
where three matrices $\Pi $, 
$\hat {U}_N$ and $\hat {V}_N$ read as 
\begin{eqnarray}  &&
\Pi =\left [\begin{array} {cc} 0&0\vspace{2mm}\\
I_{2N} &0 \end{array}  \right ]_{2(N+1)\times 2(N+1)},\ 
I_{2N}=\textrm{diag}(\underbrace{I_2,\cdots ,I_2}_{N})
=\textrm{diag}(\underbrace{1,\cdots ,1}_{2N}),
\\ &&
\hat{U}_N=\left[\begin{array} {cccc} U_0 & & & 0 \vspace{2mm}\\
U_1 & U_0 & & \vspace{2mm}\\ 
\vdots &\ddots &\ddots & \vspace{2mm}\\
U_N &\cdots &U_1&U_0 \end{array}  \right],
\ \hat{V}_N=\left[\begin{array} {cccc} V_0 & & & 0 \vspace{2mm}\\
V_1 & V_0 & & \vspace{2mm}\\ 
\vdots &\ddots &\ddots & \vspace{2mm}\\
V_N &\cdots &V_1&V_0 \end{array}  \right],
\end{eqnarray} 
with the $U_i,V_i$ being determined by
\begin{eqnarray}  && U_i
 =\left.\frac 1{i!}\frac {\part ^i U(\hat {u}_N,\hat {\lambda}_N)}
{\part \varepsilon ^i} \right|_{\varepsilon=0}
 =\left[\begin{array} {cc} 0&- \eta _i-\mu _i\vspace{2mm} \\
\delta _{i0} &0 \end{array} \right],
 \ 0\le i\le N,
\\ && V_i
 =\left.\frac 1{i!}\frac {\part ^i V(\hat {u}_N,\hat {\lambda}_N)}
{\part \varepsilon ^i} \right|_{\varepsilon=0}
 =\left[\begin{array} {cc} \eta _{ix}+\eta _{i-1,y}&
Q_i
\vspace{2mm} \\ 
 2\eta _i-4\mu _i & -\eta _{ix}-\eta _{i-1,y} \end{array} \right],
 \ 0\le i\le N,\\
&& Q_i=
- \eta _{ixx}-2\eta _{i-1,xy}-\eta _{i-2,yy}-2\sum_{k+l=i}(\eta _k\eta _l
-\mu _k\eta _l-2\mu _k\mu _l), \ 0\le i\le N,\quad  
\end{eqnarray} 
where we accept that $\eta _{-1}=\eta _{-2}=0$, and $U,V$ are defined by
(\ref{zcrepofKdV}).
Of course, we require the condition (\ref{conditionsofspectralparameters}) 
on the involved spectral parameters 
$\mu _i$, $0\le i\le N$,
in order to guarantee the equivalence between the system
(\ref{2+1KdVperturbationsystem}) and the zero curvature equation
(\ref{zcrepof2+1perturbationsystem}). 

In particular, the first-order bi-scale perturbation system
\begin{equation}\left \{\begin{array}{l}
\eta  _{0t}=\eta  _{0xxx}+6\eta  _0\eta  _{0x},\vspace{2mm}\\
\eta  _{1t}=\eta  _{1xxx}+3\eta  _{0xxy}+6(\eta  _0\eta  _{1})_x
+6\eta  _0\eta  _{0y},
\end{array}\right. \label{2+1firstorderperturbationsystemofKdV}
\end{equation}
has a local $2+1$ dimensional bi-Hamiltonian formulation  
\begin{eqnarray}  && \hat{\eta }_{1t}
 =\hat {J}_1\frac {\delta (\textrm{per}_1\tilde{H}_{1})}
{\delta \hat {\eta }_1}=
\hat {M}_1\frac {\delta (\textrm{per}_1\tilde{H}_{0})}
{\delta \Hat{ {\eta} }_1},\
\hat{\eta }_1= \left[\begin{array} {c} \eta _{0}\vspace{2mm} \\ \eta _{1}
\end{array} \right],
 \\ &&
\hat {J}_1= \left[\begin{array} {cc} 0 & \part _x \vspace{2mm} \\ 
\part _x &\part _y \end{array} \right],\ 
\hat {M}_1= \left[\begin{array}
{cc} 0&\part _x^3 +2\eta _{0x}+4\eta _0\part _x
\vspace{2mm} \\ \part _x^3 +2\eta _{0x}+4\eta _0\part _x &
3\part _x^2\part _y +2\eta _{1x}+2\eta _{0y}+4\eta _1\part _x
+4\eta _0\part _y
\end{array} \right],\qquad \quad\\ &&
\textrm{per}_1\tilde{H}_{0}=\iint \eta _0\eta _1\,dxdy,\
\textrm{per}_1\tilde{H}_{1}=\iint (\frac 12
\eta _0\eta _{1xx}+\eta _0\eta _{0xy}+\frac 12 \eta _1\eta _{0xx}
+3\eta _0^2\eta _1)\,dxdy.
 \end{eqnarray} 
Here the extended variables $\eta _0(x,y,t) $ and $
\eta _1(x,y,t)$ are taken as a potential vector $\hat {\eta }_1$.
Moreover the above Hamiltonian pair yields
a hereditary recursion operator in $2+1$ dimensions
\begin{equation}\hat {\Phi }_{1}(\hat {\eta }_1) = 
\left[\begin{array} {cc} \part _x^2 +2\eta _{0x}\part _x^{-1} +4\eta _0& 0
\vspace{2mm} \\
2\part _x\part _y-2\eta _{0x}\part _x^{-2}\part _y
+2(\eta _{1x}+\eta _{0y})\part _x^{-1}+4\eta _1
&  \part _x^2 +2\eta _{0x}\part _x^{-1} +4\eta _0
 \end{array} \right].
\end{equation} 
The system (\ref{2+1firstorderperturbationsystemofKdV}) was
furnished in \cite{MaF-PLA1996},
its Painlev\'e property and zero curvature representation
were discussed by Sakovich \cite{Sakovich-JNMP1998}, and its localized
soliton-like solutions were found in \cite{ZhouM-preprint}.
All these properties show that the system 
(\ref{2+1firstorderperturbationsystemofKdV}) is a good example of typical 
soliton equations in $2+1$ dimensions.

\section{Concluding remarks}
\setcounter{equation}{0}

We have developed a theory for constructing integrable couplings 
of soliton equations by perturbations. The symmetry problem is viewed as a 
special case of integrable couplings.
The general structures of hereditary recursion  operators, Hamiltonian
operators, symplectic operators, Hamiltonian formulations etc.
have been established under the multi-scale perturbations.
The perturbation systems have richer structures
of Lax representations and zero curvature
representations than the original systems.
For example, in the higher dimensional
cases, the involved spectral parameters $\mu _i,\ 0\le i\le N,$
may vary with respect to the spatial variables,
but they need to satisfy some conditions, for example,
 \[ \mu_{0x}=0,\ \mu _{ix}+\mu _{i-1,y}=0,\ 1\le i\le N,\]   
in the $2+1$ dimensional case of the perturbation
\[ \hat {u}_n=\sum_{i=0}^N \varepsilon ^i\eta_i (x,y,t)=
 \sum_{i=0}^N \varepsilon ^i\eta _i(x,\varepsilon x,t),\ x\in\R .
 \]
The resulting theory has been applied to the KdV soliton hierarchy
and thus various integrable couplings are presented for
each soliton equation in the KdV hierarchy.
The obtained integrable couplings of the original KdV equations 
have infinitely many commuting symmetries and conserved densities. 
Linear combinations of the KdV hierarchy containing a small
perturbation parameter may yield much more interesting integrable couplings.
For example, the KdV type systems of soliton equations 
possessing both multi-Hamiltonian formulations and two classes of
hereditary recursion operators have been presented and what's more,
local $2+1$ dimensional bi-Hamiltonian systems of the KdV type with
hereditary structures have also been constructed.  

Our success in extending the standard perturbation cases
to the non-standard cases and the higher dimensional cases are based on 
the following two simple ideas. First, we 
chose the perturbed systems as initial systems
to generate integrable couplings for given integrable systems.
The method of construction is similar to that in 
\cite{MaF-CSF1996}.
Only a slight difference is that new initial 
systems themselves involve a small perturbation parameter, but
importantly, such initial perturbed systems take effect 
in getting new integrable couplings.
In particular, our result showed that 
 the following non-standard perturbation system
\begin{equation}  \left \{ \begin{array} {l} u_t=K(u),\\ v_t=K'(u)[v]+K(u),
\end{array}   \right.  \end{equation}
keeps complete integrability. Therefore, this
also provides us with an extension
of integrable systems. Secondly, we took the multi-scale perturbations,
by which higher dimensional integrable couplings can be presented.
Indeed, the multi-scale perturbations enlarge the spatial dimensions
and keeps complete integrability of the system under study.
A concrete example of integrable couplings resulted from the multi-scale 
perturbations is the following system
\[ \left \{\begin{array}{l}
\eta  _{0t}=\eta  _{0xxx}+6\eta  _0\eta  _{0x},\vspace{2mm}\\
\eta  _{1t}=\eta  _{1xxx}+3\eta  _{0xxy}+6(\eta  _0\eta  _{1})_x
+6\eta  _0\eta  _{0y},
\end{array}\right. 
\]
which has been proved to be a local bi-Hamiltonian system.

A kind of reduction of the standard multi-scale
perturbations defined by (\ref{multiplescaleperturbationseries})
may be taken, which can be generally represented as
\[
\hat{  u}_N=\sum_{j=0}^N \varepsilon  ^{i_j}\eta _j , \
\eta _i=\eta _i(y_0, y_1, y_2,\cdots,y_r,t),\
y_j=\varepsilon ^{i_j'} x, \ t\in \R,\ x\in \R ^p,\ 0\le i\le N,
\]
where the $i_j,i_j'$ can be any two finite sets of natural numbers.
This kind of perturbations can be generated from
the standard perturbations (\ref{multiplescaleperturbationseries}),
if some dependent variables $\eta_i$ are chosen to be zero and
the other dependent variables are assumed to be independent of some dependent
variables $y_i$.
They yield more specific integrable couplings.
There is another interesting problem related integrable couplings.
Could one reduce the spatial dimensions of a given integrable system
while formulating integrable couplings? 
If the answer is yes, it is of interest to find some ways
to construct such kind of integrable couplings, i.e.,
to hunt for the second part $S(u,v)$ with $v$ being less dimensional than $u$
to constitute integrable systems with the original system
$u_t=K(u)$.

There exist some important works to deal with
asymptotic analysis and asymptotic integrability 
\cite{Wong-book1989,FokasL-PRL1996,Kodama-book1997,DegasperisP-preprint1998},
to which the study of the perturbation systems may be helpful.
It is also worthy mentioning that our $2+1$ dimensional
hereditary recursion operators,
for example, the operators defined by (\ref{2+1dimensionalrecursionoperator})
and (\ref{concreteexpressionforderivativeofhatPhiN}), 
are of the form described only by independent variables involved.
Thus they are a supplement to 
a class of recursion operators in $2+1$ dimensions
discussed by Zakharov and Konopelchenko \cite{ZakharovK-CMP1984},
and a class of the extended recursion operators in $2+1$ dimensions
including additional independent variables, introduced
by Santini and Fokas \cite{SantiniF-CMP1988} and Fokas and Santini
\cite{FokasS-CMP1988}. The other properties such 
as B\"acklund transformations, bilinear forms and soliton solutions
might be found for the resulting perturbation systems.
A remarkable Miura transformation \cite{Miura-JMP1968} 
might also be introduced for the 
$2+1$ dimensional perturbation systems (\ref{2+1KdVperturbationsystem}),
which will lead to new $2+1$ dimensional integrable systems of the MKdV type.
All these problems will be analyzed in a further publication.

\vskip 1mm
\noindent{\bf Acknowledgments:} 
The author would like to thank the City University of Hong Kong and the  
Research Grants Council of Hong Kong
for financial support. 

\setlength{\baselineskip}{15pt}

\vskip 5mm

\noindent 
DEPARTMENT OF MATHEMATICS, CITY UNIVERSITY OF HONG KONG, KOWLOON, HONG KONG

\noindent Email: mawx@math.cityu.edu.hk


\small
\begin{thebibliography}{99}

\bibitem{Fuchssteiner-book1993} B. Fuchssteiner,
{\it Coupling of completely integrable systems: the perturbation bundle},
  in: {Applications of Analytic and
  Geometric Methods to Nonlinear Differential Equations},
  (P. A. Clarkson, ed.) Kluwer, Dordrecht, 1993, pp. 125--138.
\bibitem{Ma-JPM1992} W. X. Ma, {\it Lax representations and Lax operator 
  algebras of isospectral and nonisospectral hierarchies of evolution 
  equations},
  J. Math. Phys. {\bf 33} (1992), 2464--2476.
\bibitem{Ma-JPAPLA199293} W. X. Ma, {\it An approach for 
 constructing nonisospectral hierarchies of evolution equations}, 
 J. Phys. A: Math. Gen. {\bf 25} (1992), L719--L726; 
 {\it A simple scheme for generating nonisospectral flows from zero 
  curvature representation}, Phys. Lett. A {\bf 179} (1993), 179--185.
\bibitem{LakshmananT-JMP1985} M. Lakshmanan and K. M. Tamizhmani,
{\it Lie-B\"acklund symmetries of certain nonlinear evolution equations 
  under perturbation around solutions}, 
  J. Math. Phys. {\bf 26} (1985), 1189--1200.
\bibitem{MaF-CSF1996} W. X. Ma and B. Fuchssteiner, 
  {\it Integrable theory of the perturbation equations},
  Chaos, Solitons $\&$ Fractals {\bf 7} (1996), 1227--1250.
\bibitem{Ma-Needs98}W. X. Ma, {\it Bi-Hamiltonian structures of triangular
  systems by perturbations}, preprint (1998).
\bibitem{FuchssteinerF-PD1981} B. Fuchssteiner and A. S. Fokas,
  {\it Symplectic structures, their B\"acklund transformations and 
  hereditary symmetries}, Physica D {\bf 4} (1981), 47--66.
\bibitem{Oevel-JMP1988}W. Oevel,
  {\it Dirac constraints in field theory: lifts of
  Hamiltonian systems to the
  cotangent bundle}, J. Math. Phys. {\bf 29} (1988), 210--219.
\bibitem{Fuchssteiner-NATMA1979}B. Fuchssteiner, {\it 
  Application of hereditary symmetries to nonlinear evolution equations}, 
  Nonlinear Analysis TMA {\bf 3} (1979), 849--862.
\bibitem{Fokas-SAM1987} A. S. Fokas, 
  {\it Symmetries and integrability}, 
  Studies in Appl. Math. {\bf 77} (1987), 253--299.
\bibitem{Gu-book1995}C. H. Gu, ed.
  {\it Soliton Theory and Its Applications},
  Springer-Verlag, Berlin, 1995.
\bibitem{Olver-book1986} P. J. Olver, {\it Applications of Lie Groups
  to Differential Equations}, Springer-Verlag, New York, 1986.
\bibitem{GelfandD-FAA1979}I. M. Gelfand and I. Ya. Dorfman,
  {\it Hamiltonian operators and algebraic structures related to them}, 
  Func. Anal. Appl. {\bf 13} (1979), 248--262. 
\bibitem{Dorfman-book1993} I. Ya. Dorfman, {\it Dirac Structures and 
  Integrability of Nonlinear Evolution Equations}, 
  John Wiley \& Sons, Chichester, 1993.
\bibitem{Magri-JMP1978}F. Magri,
  {\it A simple model of the integrable Hamiltonian
  equation}, J. Math. Phys. {\bf 19} (1978), 1156--1162.
\bibitem{Ma-JPA1992} W. X. Ma, {\it The algebraic structures of 
  isospectral Lax operators and applications to integrable equations},
  J. Phys. A: Math. Gen. {\bf 25} (1992), 5329--5343.
\bibitem{Miura-SIAMR1976}
  R. M. Miura, {\it The Korteweg-de Vries equation: a survey of results},
  SIAM Rev. {\bf 18} (1976), 
  412--459. 
\bibitem{AblowitzS-book1981} M. J. Ablowitz and
  H. Segur, {\it Solitons and the Inverse Scattering Transform},
  SIAM, Philadelphia, 1981.
\bibitem{TamizhmaniL-JPA1983} K. M. Tamizhmani and M. Lakshmanan,
  {\it Complete integrability of the Korteweg-de Vries
  equation under perturbation around its solution: 
  Lie-B\"acklund symmetry approach}, 
  J. Phys. A: Math. Gen. {\bf 16} (1983), 3773--3782.
\bibitem{MaF-PLA1996} W. X. Ma and B. Fuchssteiner, 
  {\it The bi-Hamiltonian structure of 
  the perturbation equations of the KdV hierarchy},
  Phys. Lett. A {\bf 213} (1996), 49--55.
\bibitem{Sakovich-JNMP1998}S. Yu. Sakovich,
  {\it On integrability of a $(2+1)$-dimensional perturbed KdV equation},
  J. Nonlinear Math. Phys. {\bf 5} (1998), 230--233.
\bibitem{ZhouM-preprint} Z. X. Zhou and W. X. Ma,
  {\it Darboux transformations for a 2+1 dimensional matrix 
  Gelfand-Dickey system}, preprint, 1998.
\bibitem{Wong-book1989}R. Wong, {\it Asymptotic Approximations of Integrals}, 
  Computer Science and Scientific Computing, Academic Press, Boston, 1989.
\bibitem{FokasL-PRL1996}A. S. Fokas and Q. M. Liu, 
  {\it Asymptotic integrability
  of water waves}, Phys. Rev. Lett. {\bf 77} (1996),  
  2347--2351.
\bibitem{Kodama-book1997}Y. Kadama and A. V. Mikhailov,
  {\it Obstacles to Asymptotic Integrability}, in: Algebraic Aspects of
  Integrable Systems: in memory of Irene Dorfman, 
  Progr. Nonlinear Differential Equations Appl. 26 
  (A. S. Fokas
  and I. M. Gelfand. eds.), Birkh\"auser, Boston, 1997, pp. 173--204.
\bibitem{DegasperisP-preprint1998}A. Degasperis and M. Procesi,
  {\it A test of asymptotic integrability of $1+1$ wave equations},
  preprint, 1998.
\bibitem{ZakharovK-CMP1984}
  V. E. Zakharov and B. G. Konopelchenko, {\it 
  On the theory of recursion operator}, Comm. Math.
  Phys. {\bf 94} (1984), 
  483--509.
\bibitem{SantiniF-CMP1988} P. M.
  Santini and A. S. Fokas, {\it Recursion operators and 
  bi-Hamiltonian structures in multidimensions. I.},
  Comm. Math. Phys. {\bf 115} (1988), 
  375--419. 
\bibitem{FokasS-CMP1988}
  A. S. Fokas and P. M. Santini, 
  {\it Recursion operators and bi-Hamiltonian structures 
  in multidimensions. II.},
  Comm. Math. Phys. {\bf 116} (1988), 
  449--474. 
\bibitem{Miura-JMP1968}R. M. Miura, 
  {\it Korteweg-de Vries equation and generalizations. 
   I. A remarkable explicit nonlinear transformation}, 
   J. Math. Phys. {\bf 9} (1968), 1202--1204.

\end{thebibliography}
\end{document}